\documentclass[useAMS,usenatbib]{mnras}
\usepackage{myaasmacros}
\usepackage{graphicx}
\usepackage{ulem}
\usepackage[table]{xcolor}
\usepackage{amsmath,amssymb} 
\usepackage{pdflscape}
\usepackage{geometry}

\def\ltsima{$\; \buildrel < \over \sim \;$}
\def\simlt{\lower.5ex\hbox{\ltsima}}   
\def\gtsima{$\; \buildrel > \over \sim \;$}
\def\simgt{\lower.5ex\hbox{\gtsima}}

\def\coreNFW{{\sc coreNFW}}
\def\NFW{{\sc NFW}}
\def\EMCEE{{\sc emcee}}
\def\Barolo{{\sc $^{3\rm D}$Barolo}}
\def\MstarM{$M_*-M_{200}$}
\def\MstarMrot{$\left.M_*-M_{200}\right|_{\rm rot}$}
\def\MstarMabund{$\left.M_*-M_{200}\right|_{\rm abund}$}

\usepackage{array}
\newcolumntype{L}[1]{>{\raggedright\let\newline\\\arraybackslash\hspace{0pt}}m{#1}}
\newcolumntype{C}[1]{>{\centering\let\newline\\\arraybackslash\hspace{0pt}}m{#1}}
\newcolumntype{R}[1]{>{\raggedleft\let\newline\\\arraybackslash\hspace{0pt}}m{#1}}

\title[The stellar mass-halo mass relation of isolated field dwarfs]{The stellar mass-halo mass relation of isolated field dwarfs: a critical test of $\Lambda$CDM at the edge of galaxy formation}

\author[J. I. Read et al.]{J. I. Read$^{1}$\thanks{E-mail: justin.inglis.read@gmail.com}, G. Iorio$^{2,3}$, O. Agertz$^1$, F. Fraternali$^{2,4}$\\
$^1${\small Department of Physics, University of Surrey, Guildford, GU2 7XH, Surrey, UK}\\
$^2${\small Dipartimento di Fisica e Astronomia, Universit\`a di Bologna, Viale Berti Pichat 6/2, I-40127, Bologna, Italy}\\
$^3${\small INAF -- Osservatorio Astronomico di Bologna, via Ranzani 1, I-40127, Bologna, Italy}\\
$^4${\small Kapteyn Astronomical Institute, University of Groningen, Landleven 12, 9747 AD Groningen, The Netherlands}\\
}

\begin{document}

\maketitle

\begin{abstract}
We fit the rotation curves of isolated dwarf galaxies to directly measure the stellar mass-halo mass relation (\MstarM) over the mass range $5 \times 10^5 \simlt M_{*}/{\rm M}_\odot \simlt 10^{8}$. By accounting for cusp-core transformations due to stellar feedback, we find a monotonic relation with little scatter. Such monotonicity implies that abundance matching should yield a similar \MstarM\ if the cosmological model is correct. Using the `field galaxy' stellar mass function from the Sloan Digital Sky Survey (SDSS) and the halo mass function from the $\Lambda$ Cold Dark Matter Bolshoi simulation, we find remarkable agreement between the two. This holds down to $M_{200} \sim 5 \times 10^9$\,M$_\odot$, and to $M_{200} \sim 5 \times 10^8$\,M$_\odot$ if we assume a power law extrapolation of the SDSS stellar mass function below $M_* \sim 10^7$\,M$_\odot$. 

However, if instead of SDSS we use the stellar mass function of nearby galaxy groups, then the agreement is poor. This occurs because the group stellar mass function is shallower than that of the field below $M_* \sim 10^9$\,M$_\odot$, recovering the familiar `missing satellites' and `too big to fail' problems. Our result demonstrates that both problems are confined to group environments and must, therefore, owe to `galaxy formation physics' rather than exotic cosmology. 

Finally, we repeat our analysis for a $\Lambda$ Warm Dark Matter cosmology, finding that it fails at 68\% confidence for a thermal relic mass of $m_{\rm WDM} < 1.25$\,keV, and $m_{\rm WDM} < 2$\,keV if we use the power law extrapolation of SDSS. We conclude by making a number of predictions for future surveys based on these results.
\end{abstract}


\section{Introduction}\label{sec:intro}
The standard $\Lambda$ Cold Dark Matter ($\Lambda$CDM) cosmological model gives an excellent description of the growth of structure in the Universe, matching the observed temperature fluctuations in the cosmic microwave background radiation \citep[e.g.][]{1992ApJ...396L...1S,2013arXiv1303.5076P}; the growth of large scale structure \citep[e.g.][]{2006Natur.440.1137S}; the clustering of galaxies \citep{2016MNRAS.455.4301C}; large scale weak lensing distortions \citep[e.g.][]{1991MNRAS.251..600B,2014MNRAS.441.2725F}; baryon acoustic oscillations \citep[e.g.][]{2003ApJ...594..665B,2005ApJ...633..560E,2013AJ....145...10D}; and the flux power spectrum of quasar absorption lines \citep[e.g.][]{1998ApJ...495...44C,2015arXiv151201981B}. However, over the past two decades there have been persistent tensions claimed on small scales inside galaxy groups and individual galaxies. These include:

\begin{enumerate}
\item {\it The `missing satellites' problem}: Pure dark matter cosmological simulations of structure formation predict that thousands of bound dark matter halos should orbit the Milky Way and Andromeda, yet only a few tens of visible satellites have been observed to date \citep[e.g.][]{1999ApJ...522...82K,1999ApJ...524L..19M,2012AJ....144....4M}. 

\item {\it The `cusp-core' problem}: These same simulations predict that the dark matter density distribution within galaxies should be self-similar and well fit at the ${\sim}10$\% level by the `NFW' profile (\citealt{1996ApJ...462..563N}):

\begin{equation} 
\rho_{\rm NFW}(r) = \rho_0 \left(\frac{r}{r_s}\right)^{-1}\left(1 + \frac{r}{r_s}\right)^{-2}
\label{eqn:rhoNFW}
\end{equation}
where the central density $\rho_0$ and scale length $r_s$ are given by: 
\begin{equation} 
\rho_0 = \rho_{\rm crit} \Delta c^3 g_c / 3 \,\,\,\, ; \,\,\,\, r_s = r_{200} / c
\end{equation}
\begin{equation}
g_c = \frac{1}{{\rm log}\left(1+c\right)-\frac{c}{1+c}} \,\,\,\, ; \,\,\,\, 
r_{200} = \left[\frac{3}{4} M_{200} \frac{1}{\pi \Delta \rho_{\rm crit}}\right]^{1/3}
\label{eqn:gcr200}
\end{equation} 
$c$ is the dimensionless `concentration parameter'; $\Delta = 200$ is the over-density parameter; $\rho_{\rm crit}$ is the critical density of the Universe today; $r_{200}$ is the `virial' radius at which the mean enclosed density is $\Delta \times \rho_{\rm crit}$; and $M_{200}$ is the `virial' mass within $r_{200}$.

For over two decades now, the rotation curves of small dwarf and low surface brightness galaxies have favoured a central constant density core over the `cuspy' \NFW\ profile described above \citep[e.g.][]{1994ApJ...427L...1F,1994Natur.370..629M,2002A&A...385..816D,2011MNRAS.414.3617K,2011AJ....142...24O,2013MNRAS.433.2314H}.

\item {\it The `too big to fail' problem (TBTF)}: The central velocity dispersion of Local Group dwarfs appears to be too low to be consistent with the most massive subhalos in $\Lambda$CDM \citep{2006MNRAS.tmp..153R,2011MNRAS.415L..40B}.

\end{enumerate} 
The above puzzles could be hinting at physics beyond $\Lambda$CDM, for example exotic inflation models \citep[e.g.][]{2002PhRvD..66d3003Z}, or exotic dark matter models \citep[e.g.][]{1994Natur.370..629M,2013MNRAS.430...81R,2014arXiv1412.1477E}. However, it is important to emphasise that all of these puzzles arise from a comparison between the observed Universe and a model $\Lambda$CDM universe entirely devoid of stars and gas (that we shall refer to from here on as `baryons'; e.g. see the discussion in \citealt{2014Natur.506..171P} and \citealt{2014JPhG...41f3101R}). Semi-analytic models make some attempt to improve on this by painting stars onto pure dark matter simulations \citep[e.g.][]{2006RPPh...69.3101B}. However, implicit in such analyses is an assumption that the distribution of dark matter is unaltered by the process of galaxy formation. It is becoming increasingly likely that this assumption is poor, especially within group environments and on the scale of tiny dwarf galaxies. 

\citet{1996MNRAS.283L..72N} were the first to suggest that dark matter could be collisionlessly heated by impulsive gas mass loss driven by supernova explosions. They found that, for reasonable initial conditions corresponding to isolated dwarf galaxies, the effect is small \citep[see also][]{2002MNRAS.333..299G}. However, \citet{2005MNRAS.356..107R} showed that the effect can be significant if star formation proceeds in repeated bursts, gradually grinding a dark matter cusp down to a core. There is mounting observational evidence for such bursty star formation \citep{2012ApJ...750...33L,2013MNRAS.429.3068T,2012ApJ...744...44W,2014MNRAS.441.2717K,2015MNRAS.450.3886M}, while the physics of such `cusp-core transformations' is now well-understood (\citealt{2012MNRAS.421.3464P,2015arXiv150207356P}, and for a review see \citealt{2014Natur.506..171P}). The latest numerical simulations that resolve the effect of individual supernovae explosions are substantially more predictive \citep[e.g.][hereafter R16a]{2015arXiv150202036O,2015MNRAS.454.2981C,2015arXiv150804143R}; these demonstrate that dark matter cores are an unavoidable prediction of $\Lambda$CDM (with baryons) for all low mass dwarf galaxies, so long as star formation proceeds for long enough\footnote{Two recent studies have claimed that dark matter cores do not form at any mass scale \citep{2016MNRAS.457.1931S,2016MNRAS.458.1559Z}. However, both of these used simulations with a `cooling floor' of $10^4$\,K, meaning that they are unable to resolve the clumpy interstellar medium. Resolving this is crucial for exciting cusp-core transformations, as explained in \citet{2012MNRAS.421.3464P}.}.

However, there remains a debate in the literature over the efficiency of star formation in low mass halos. \citet{2014MNRAS.441.2986D}, \citet{2015MNRAS.454.2981C} and \citet{2015arXiv150703590T} find insufficient star formation to excite cusp-core transformations below $M_{200} \sim 10^{10}$\,M$_\odot$;  \citet{2014ApJ...789L..17M} find that core formation proceeds in $M_{200} \sim 10^9$\,M$_\odot$ dwarfs; and R16a find that core formation proceeds `all the way down' to halo masses ${\sim}10^8$\,M$_\odot$. These differences owe in part to resolution. R16a have a typical spatial resolution of 4\,pc for their isolated dwarfs, with a stellar and dark matter particle mass resolution of ${\sim}250$\,M$_\odot$. This allows them to resolve the ${\simlt}500$\,pc size cores that form in their $M_{200} \simlt 10^9$\,M$_\odot$ dwarfs. Such small cores cannot be captured by the \citet{2014MNRAS.441.2986D} and \citet{2015arXiv150703590T} simulations that have a spatial resolution of ${\sim}80-100$\,pc. However, \citet{2015MNRAS.454.2981C} have a spatial resolution of ${\sim}30$\,pc for their $10^9$\,M$_\odot$ dwarf, yet they find that no significant dark matter core forms. This owes to a second key difference between these studies: the treatment of reionisation. In R16a, reionisation is not modelled and so star formation is allowed to proceed unhindered at very low halo mass. In all of the other studies, some model of reionisation heating is included. But the mass scale at which reionisation begins to suppress star formation, $M_{\rm reion}$, remains controversial. Some recent simulations favour a high $M_{\rm reion} \sim 10^{10}$\,M$_\odot$ \citep[e.g.][]{2015MNRAS.454.2981C,2015arXiv150703590T}, while others favour a much lower $M_{\rm reion} \sim 5 \times 10^8$\,M$_\odot$ \citep{2014ApJ...793...30G}, consistent with the assumption of no reionisation in R16a. Observationally, the continuous low star formation rate of nearby dwarf irregular galaxies (dIrrs) appears to favour a low $M_{\rm reion}$ \citep[][and see the discussion in R16a]{2009MNRAS.392L..45R,2012ApJ...748...88W}. We will discuss $M_{\rm reion}$ further in \S\ref{sec:reionisation}.

Despite the differences in $M_{\rm reion}$, all of the above studies find that when dark matter cores do form, they are of size $\sim$ the projected half stellar mass radius ($R_{1/2}$). Such cores are dynamically important by construction because they alter the dark matter distribution precisely where we can hope to measure it using stellar kinematics (R16a). They also have important effects beyond just the internal structure of galaxies. Cored dwarfs are much more susceptible to tidal shocking and stripping on infall to a larger host galaxy \citep[e.g.][]{2006MNRAS.tmp..153R,2010MNRAS.406.1290P,2013ApJ...765...22B}. This aids in the morphological transformation of dwarfs from discs to spheroids \citep{2001ApJ...547L.123M,2012ApJ...751L..15L,2013ApJ...764L..29K}; and physically reshapes the dark matter halo mass function within groups \citep[][and see the discussion in R16a]{2010MNRAS.406.1290P,2012ApJ...761...71Z,2013ApJ...765...22B,2016arXiv160205957W}.

Using simulations of isolated dwarfs at a spatial and mass resolution of ${\sim}4$\,pc and ${\sim}250$\,M$_\odot$, respectively, R16a derived a new `\coreNFW' fitting function that describes cusp-core transformations in $\Lambda$CDM over the mass range $10^8 \simlt M_{200}/{\rm M}_\odot \simlt 10^{10}$ (see equation \ref{eqn:coreNFW}). In \citet{2016arXiv160105821R} (hereafter R16b), we showed that this gives a remarkable match to the rotation curves 
of four isolated dwarf irregular galaxies, using just two free fitting parameters: $M_{200}$ and $c$ (that take on the same meaning as in equation \ref{eqn:rhoNFW} for the \NFW\ profile). In particular, using mock data, we demonstrated that if the data are good enough (i.e. if the dwarfs are not face-on; starbursting; and/or of uncertain distance) then we are able to successfully measure both $M_{200}$ and $c$ within our quoted uncertainties.

In this paper, we apply the rotation curve fitting method described in R16b to 19 isolated dwarf irregulars (dIrrs) to measure the stellar mass-halo mass relation \MstarMrot\ over the stellar mass range $5 \times 10^5 \simlt M_{*}/{\rm M}_\odot \simlt 10^{8}$. We then compare this with the stellar mass-halo mass relation obtained from `abundance matching', \MstarMabund, to arrive at a comparatively clean test of our current cosmological model. 

This paper is organised as follows. In \S\ref{sec:newprobe}, we show how the comparison between \MstarMrot\ and \MstarMabund\ constitutes a rather clean cosmological probe at the edge of galaxy formation. In \S\ref{sec:data}, we describe our data compilation of rotation curves, stellar masses, and stellar mass functions. In \S\ref{sec:rotmethod}, we briefly review our rotation curve fitting method that is described and tested in detail in R16b. In \S\ref{sec:results}, we present the results from applying our rotation curve fitting method to 19 isolated dwarf irregular galaxies in the field (the individual fits and fitted parameters are reported in Table \ref{tab:data} and Appendix \ref{app:rotcurves}). In \S\ref{sec:discussion}, we discuss the implications of our results and their relation to previous works in the literature. Finally, in \S\ref{sec:conclusions} we present our conclusions.

\section{A clean cosmological probe at the edge of galaxy formation}\label{sec:newprobe}

In this paper, we test cosmological models by comparing the stellar mass-halo mass relation derived from galaxy rotation curves (\MstarMrot) with the mean stellar mass-halo mass relation derived from `abundance matching' (\MstarMabund). The idea in itself is not new. For example, \citet{2010ApJ...710..903M} compare \MstarMabund\ in $\Lambda$CDM with the stellar mass-halo mass relation derived from galaxy-galaxy lensing, finding good agreement. However, most studies to date have focussed on the high mass end of this relation where the differences between $\Lambda$CDM and alternative cosmologies like $\Lambda$ Warm Dark Matter ($\Lambda$WDM) are small (e.g. \citealt{2009MNRAS.394..929C,2015arXiv150200313S}; and see Figure \ref{fig:carina_plot}). More recently, \citet{2016arXiv160505326P} and \citet{2016arXiv160505971K} have used the baryon-influenced mass models from \citet{2014MNRAS.441.2986D} to fit rotation curves and measure $M_{200}$ and $c$ for a large sample of dwarfs, comparing their results with abundance matching predictions. We will compare and contrast our analysis with these studies in \S\ref{sec:discussion}. However, what is new to this paper are the following key ingredients: (i) we focus on building a particularly clean sample of rotation curves, derived in a consistent manner and with a state-of-the-art technique \Barolo\ (\citealt{2015MNRAS.451.3021D}; \citealt{2016arXiv161103865I}); (ii) we perform our comparison at $M_* \simlt 10^8$\,M$_\odot$, maximising the constraints on cosmological models; and (iii) we make use of a new {\it predictive} \coreNFW\ profile for the dark matter distribution on these mass scales that accounts for cusp-core transformations due to stellar feedback (R16a,b). In the remainder of this section, we discuss in detail how our cosmological test works and why it is particularly clean.

Classical abundance matching relies on a key assumption that galaxy stellar masses are monotonically related to dark matter halo masses \citep{2004MNRAS.353..189V}. Armed with this, galaxies are mapped to dark matter halos of the same cumulative number density, providing a statistical estimate of \MstarMabund\ for a given cosmological model. Thus, by comparing this \MstarMabund\ with \MstarMrot, we arrive at a comparatively clean cosmological probe of structure formation on small scales. The probe is clean because it relies only on the following theoretical ingredients:

\begin{enumerate}
\item A monotonic relation between stellar mass and halo mass. We will directly test this with our measurement of \MstarMrot\ in \S\ref{sec:mstarmrot}.
\item The dark matter halo mass function. This is readily calculated for a given cosmological model using cosmological simulations \citep[e.g.][]{2011EPJP..126...55D,2011ApJ...740..102K}.
\item A robust prediction of the internal dark matter distribution in dwarf irregular galaxies $\rho_{\rm dm}(r)$, for a given cosmological model. This is required in order to measure $M_{200}$ from rotation curve data to obtain \MstarMrot. In \S\ref{sec:exrots} we show that, while our \coreNFW\ dark matter density profile gives a significantly better fit to our sample of rotation curves than the NFW profile, we are not particularly sensitive to this choice so long as $\rho_{\rm dm} \rightarrow \rho_{\rm NFW}$ (see equation \ref{eqn:rhoNFW}) for $r > R_{1/2}$.
\end{enumerate}
Armed with the above theory ingredients, our probe relies solely on observational data: rotation curves for dwarf galaxies with well measured inclination and distance, and no evidence of a recent starburst (see R16b); stellar masses derived from SED model fitting\footnotemark; and a good measure of the stellar mass function of galaxies.
\footnotetext{
Note that such stellar masses are theoretically derived quantities, not directly measured from the data. However, this critique applies equally to the stellar masses derived for \MstarMrot\ and \MstarMabund. As such, the {\it comparison} between these two should not be sensitive to the details of our stellar mass modelling, so long as we are consistent.
}

For our abundance matching, we use as default the stellar mass function from SDSS that reaches down to $M_* \sim 10^7$\,M$_\odot$ \citep{2005ApJ...631..208B,2008MNRAS.388..945B,2010ApJ...717..379B,2013ApJ...770...57B}; and the halo mass function from the $\Lambda$CDM `Bolshoi' simulation that is accurate to $M_{200} \sim 10^{10}$\,M$_\odot$ (\citealt{2011ApJ...740..102K}; the cosmological parameters assumed by this simulation are reported in Table \ref{tab:cosmopars}). Below these mass scales, we use power law extrapolations. We compare the SDSS stellar mass function to those derived in \citet{2005RSPTA.363.2693R} (hereafter RT05); GAMA \citep{2012MNRAS.421..621B}; and zCOSMOS \citep{2012A&A...538A.104G} in Figure \ref{fig:stellar_massfunc_compare}. The survey data are described in \S\ref{sec:stellarmassfunc}, while we explore reasons for their different faint end slopes in \S\ref{sec:stellarmfuncs}.

\begin{table}
\begin{center}
\begin{tabular}{l|l}
{\bf Cosmological Parameter} & {\bf Value} \\
\hline
\hline
Hubble $h$ & 0.7 \\
$\Omega_{\rm M}$ & 0.27 \\
$\Omega_{\Lambda}$ & 0.73 \\
Tilt $n$ & 0.95 \\
$\sigma_8$ & 0.82 \\
\hline
\end{tabular}
\caption{Cosmological parameters assumed in this work. From top to bottom, these are: the Hubble parameter; the ratio of matter and dark energy density to the critical density; the `tilt' of the power spectrum; and the amplitude of the power spectrum on a scale of 8$h^{-1}$\,Mpc \citep[see e.g.][for a full definition of these]{1999coph.book.....P}. These parameters are chosen to match those used in the Bolshoi simulation \citep{2011ApJ...740..102K} and give a good description of the latest cosmological data (see the discussion in \citealt{2011ApJ...740..102K}).}
\label{tab:cosmopars}
\end{center}
\end{table}

In addition to testing a $\Lambda$CDM cosmology, we explore an effective $\Lambda$ `warm' dark matter cosmology ($\Lambda$WDM). This corresponds to a dark matter particle that is relativistic for some time after decoupling in the early Universe, leading to a suppression in the growth of structure on small scales and at early times \citep[e.g.][]{2001ApJ...556...93B,2001ApJ...559..516A}. We describe this model in detail in \S\ref{sec:wdm}.

\section{The data}\label{sec:data}

\subsection{The rotation curve sample}

We compile HI data for 19 isolated dwarf irregular galaxies over the mass range $5 \times 10^5 \simlt M_{*}/{\rm M}_\odot \simlt 10^{8}$ from \citet{2003MNRAS.340...12W} and \citet{2015AJ....149..180O}; and stellar mass and surface density data from \citet{2012AJ....143...47Z}. Our sample selection, that primarily comprises a subset of Little THINGS galaxies, is discussed in detail in \citet{2016arXiv161103865I} and R16b. We exclude galaxies that are known to have very low inclination (for which the rotation curve extraction can become biased; R16b); four Blue Compact Dwarfs; and any galaxy for which there is no published stellar mass profile. This leaves about half of the full Little THINGS sample. \citet{2016arXiv161103865I} show that this subset is representative of the full Little THINGS survey in terms of its distribution of distances, absolute magnitudes, star formation rate densities, and baryonic masses.

In addition, we include two galaxies which do not have gaseous rotation curves: the isolated dwarf irregular Leo T and the Milky Way dwarf spheroidal galaxy Carina. We estimate $M_{200}$ for Leo T by direct comparison to the simulations in R16a. There, we showed that Leo T gave a poor match to our $M_{200} = 10^8$\,M$_\odot$ and $M_{200} = 10^9$\,M$_\odot$ simulations, but an excellent match to the photometric light profile; star formation history; stellar metallicity distribution function; and star/gas kinematics of our $M_{200} = 5 \times 10^8$\,M$_\odot$ simulation. From this comparison, we estimate $M_{200,{\rm Leo T}} = 3.5 - 7.5 \times 10^8$\,M$_\odot$ (see Table \ref{tab:data}). (A similar analysis for the Aquarius dwarf yields a mass $M_{200} \sim 10^9$\,M$_\odot$ in good agreement with its rotation curve derived value; see Table \ref{tab:data}.) For Carina, we use the pre-infall `tidal mass estimate' from \citet{2015NatCo...6E7599U}. This is derived by directly fitting $N$-body simulations of Carina tidally disrupting in the halo of the Milky Way to data for the positions and velocities of `extra-tidal' stars reported in \citet{2006ApJ...649..201M}. Leo T is interesting because it is the lowest mass dwarf discovered to date with ongoing star formation \citep{2008MNRAS.384..535R}. In R16a, we argued that its lack of a visible HI rotation curve owes to it having a low inclination ($i < 20^\circ$). Carina is interesting because, despite its close proximity to the Milky Way, it has continued to form stars for almost a Hubble time (though with notable bursts; \citealt{2014A&A...572A..10D}). We use Carina to discuss at what mean orbital distance from the Milky Way environmental effects start to play an important role, driving scatter in the \MstarMrot\ relation (\S\ref{sec:discussion}).

All of the data are summarised in Table \ref{tab:data}, including our derived model fitting parameters. We describe our methodology for extracting the rotation curves from the HI datacubes and fitting model rotation curves in \S\ref{sec:rotmethod}.

\subsection{The stellar mass functions}\label{sec:stellarmassfunc}

\begin{figure}
\begin{center}
\includegraphics[width=0.49\textwidth]{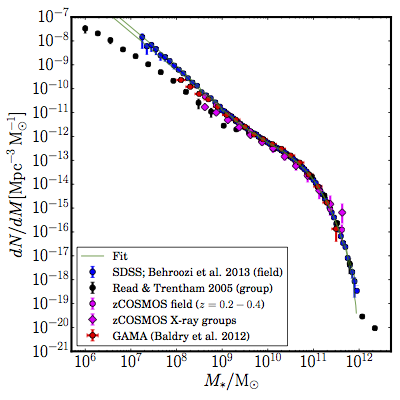}
\caption{Galaxy stellar mass functions compiled from the literature. The blue data points show the stellar mass function from SDSS \citep{2013ApJ...770...57B}; the black data points show the group stellar mass function from RT05; the red data points show the stellar mass function from GAMA \citep{2012MNRAS.421..621B}; and the magenta data points show the stellar mass functions from zCOSMOS field galaxies over the redshift range $z=0.2-0.4$ (circles) and from X-ray selected groups (diamonds; \citealt{2012A&A...538A.104G} and see \S\ref{sec:data} for further details). The green tracks show a non-parametric fit to the SDSS stellar mass function, where the upper and lower tracks encompass the 68\% confidence intervals of the data. Below $M_* = 10^8$\,M$_\odot$ we assume a power law with logarithmic slope $\alpha = 1.6$.}
\label{fig:stellar_massfunc_compare} 
\end{center}
\end{figure}

We take the SDSS stellar mass function from \citet{2013ApJ...770...57B}, which was originally obtained by   \citet{2008MNRAS.388..945B}. The uncertainty on the stellar mass is comparable to our assumed uncertainty for the isolated dwarf galaxy sample described above of ${\sim}25$\% \citep{2015AJ....149..180O}, making the comparison between the stellar masses in \citet{2013ApJ...770...57B} and those taken from \citet{2012AJ....143...47Z} entirely reasonable. 

In Figure \ref{fig:stellar_massfunc_compare}, we compare the SDSS stellar mass function (blue data points) with those derived by \citet{2005RSPTA.363.2693R} (black data points; hereafter RT05); GAMA \citep[red data points;][]{2012MNRAS.421..621B}; and zCOSMOS \citep[magenta data points;][]{2012A&A...538A.104G}. 

The SDSS stellar mass function is derived from the \citet{2005ApJ...631..208B} survey of low luminosity galaxies \citep{2008MNRAS.388..945B}. This is complete to a stellar mass of $M_* \sim 2 \times 10^7$\,M$_\odot$ over a volume of ${\sim}2 \times 10^6\,{\rm Mpc}^3$. The GAMA stellar mass function is derived from about a tenth of the SDSS survey volume (${\sim}2 \times 10^5\,{\rm Mpc}^3$), and is complete to a stellar mass of $M_* \sim 10^8$\,M$_\odot$. Due to its smaller survey volume, its stellar mass function is more prone to cosmic variance \citep{2005ApJ...631..208B,2011MNRAS.413..971D}. The zCOSMOS survey covers a small 1.7\,deg$^2$ patch of the sky, but to much higher redshift \citep{2007ApJS..172...70L}. Here, we use the lowest redshift range $0.2 < z < 0.4$ that corresponds to a volume similar to that of the GAMA survey (${\sim}1.8 \times 10^5\,{\rm Mpc}^3$), complete down to a stellar mass of $M_* \sim 4 \times 10^8$\,M$_\odot$ \citep{2012A&A...538A.104G}. The full zCOSMOS sample is split into a `field' population (magenta circles) and X-ray selected groups (magenta diamonds), both over the redshift range $0.2 < z < 0.4$. Since group environments are more dense on average, we renormalise the X-ray selected groups from \citet{2012A&A...538A.104G} to match SDSS at $M_* = 10^{10}$\,M$_\odot$. Finally, we consider the stellar mass function from RT05. At $M_* \simgt 10^9$\,M$_\odot$, this is taken from SDSS \citep{2001AJ....121.2358B,2003ApJ...592..819B}; at lower stellar mass it comes from the \citet{2002MNRAS.335..712T} catalogue of five nearby groups, including the Local Group (see \citet{2005MNRAS.tmp...52T} for details of how these surveys are sewn together). The \citet{2002MNRAS.335..712T} group catalogue is derived from deep mosaic surveys that are complete to a stellar mass of $M_* \sim 10^6$\,M$_\odot$, but cover a tiny volume as compared to SDSS of just ${\sim}5$\,Mpc$^3$.

As can be seen in Figure \ref{fig:stellar_massfunc_compare}, all of these different stellar mass functions agree within their uncertainties above $M_* \sim 10^9$\,M$_\odot$. However, at lower stellar masses there is a striking divergence between all of them bar SDSS and the zCOSMOS field stellar mass function that are in good agreement. We discuss this further, and the possible reasons for it, in \S\ref{sec:stellarmfuncs}. 

\vspace{-3mm}
\section{Extracting \& modelling dwarf galaxy rotation curves}\label{sec:rotmethod}

\subsection{Extracting rotation curves from HI data cubes}\label{sec:barolo}

Our rotation curves are derived from HI datacubes \citep{2003MNRAS.340...12W,2015AJ....149..180O} using the publicly available software \Barolo\ \citep{2015MNRAS.451.3021D}. \Barolo\ fits tilted-ring models directly to the datacube by building artificial 3D data and minimising the residuals, without explicitly extracting velocity fields (as in e.g. \citealt{2015AJ....149..180O}). This ensures full control of the observational effects and, in particular, a proper account of beam smearing that can strongly affect the derivation of the rotation velocities in the inner regions of dwarf galaxies \citep[see e.g.][]{1999PhDT........27S}. \Barolo\ was extensively tested on mock data in R16b and shown to give an excellent recovery of the rotation curve so long as the best fit inclination angle was $i_{\rm fit} > 40^\circ$. The final rotation curves were corrected for asymmetric drift, as described in R16b and \citet{2016arXiv161103865I}. The detailed description of the data analysis, including comments on individual galaxies, are presented in those papers.

\subsection{The mass model}\label{sec:massmodel}

We use the same mass model as described in detail in R16b. Briefly, we decompose the circular speed curve into contributions from stars, gas and dark matter: 

\begin{equation}
v_c^2 = v_*^2 + v_{\rm gas}^2 + v_{\rm dm}^2
\end{equation}
where $v_*$ and $v_{\rm gas}$ are the contributions from stars and gas, respectively, and $v_{\rm dm}$ is the dark matter contribution. We assume that both the stars and gas are well-represented by exponential discs: 

\begin{equation} 
v_{*/{\rm gas}}^2 = \frac{2 G M_{*/{\rm gas}}}{R_{*/{\rm gas}}} y^2 \left[I_0(y) K_0(y) - I_1(y) K_1(y)\right]
\label{eqn:vcstargas}
\end{equation}
where $M_{*/{\rm gas}}$ is the mass of the star/gas disc, respectively; $R_{*/{\rm gas}}$ is the exponential scale length; $y = R/R_{*/{\rm gas}}$ is a dimensionless radius parameter; and $I_0, I_1, K_0$ and $K_1$ are Bessel functions \citep{1987gady.book.....B}. We fix the values of $R_*$ and $R_{\rm gas}$ in advance of running our Markov Chain Monte Carlo (MCMC) models (see \S\ref{sec:mcmc}). All values used are reported in Table \ref{tab:data}.

For the dark matter profile, we use the \coreNFW\ profile from R16a: 

\begin{equation}
M_{\rm cNFW}(<r) = M_{\rm NFW}(<r) f^n
\label{eqn:coreNFW}
\end{equation}
where $M_{\rm NFW}(<r)$ is the usual NFW enclosed mass profile \citep{1996ApJ...462..563N}:

\begin{equation} 
M_{\rm NFW}(<r) = M_{200} g_c \left[\ln\left(1+\frac{r}{r_s}\right) - \frac{r}{r_s}\left(1 + \frac{r}{r_s}\right)^{-1}\right]
\label{eqn:MNFW}
\end{equation}
and $M_{200}$; $c$; $r_s$; $g_c$; $\rho_{\rm crit}$; and $\Delta = 200$ are as in equation \ref{eqn:rhoNFW}.

The function $f^n$ generates a shallower profile below a core radius $r_c$: 

\begin{equation} 
f^n = \left[\tanh\left(\frac{r}{r_c}\right)\right]^n
\end{equation}
where the parameter $0 < n \le 1$ controls how shallow the core becomes ($n=0$ corresponds to no core; $n=1$ to complete core formation). The parameter $n$ is tied to the total star formation time\footnote{More precisely, the total {\it duration} of star formation, not to be confused with the star formation depletion timescale $t_{\rm dep}=\Sigma_{\rm gas}/\Sigma_{\rm SFR}$ \citep[e.g.][]{Bigiel2011}.} $t_{\rm SF}$:
\begin{equation} 
n = \tanh(q) \,\,\,\, ; \,\,\,\, q = \kappa \frac{t_{\rm SF}}{t_{\rm dyn}}
\end{equation} 
where $t_{\rm dyn}$ is the circular orbit time at the \NFW\ profile scale radius $r_s$:
\begin{equation} 
t_{\rm dyn} = 2\pi \sqrt{\frac{r_s^3}{G M_{\rm NFW}(r_s)}}
\end{equation}
and $\kappa = 0.04$ is a fitting parameter (see R16a). For the isolated dwarfs that we consider here, we assume $t_{\rm SF} = 14$\,Gyrs such that they have formed stars continuously for a Hubble time. For this value of $t_{\rm SF}$, $n \sim 1$ and we expect the dwarfs to be maximally cored.

The core size is set by the projected half stellar mass radius of the stars $R_{1/2}$: 

\begin{equation} 
r_c = \eta R_{1/2}
\label{eqn:etarc}
\end{equation}
where, for an exponential disc, $R_{1/2} = 1.68 R_*$. By default, we assume that the dimensionless core size parameter, $\eta = 1.75$ since this gives the best match to the simulations in R16a. However, as discussed in R16a, there could be some scatter in $\eta$ due to varying halo spin, concentration parameter and/or halo assembly history. We explore our sensitivity to $\eta$ in Appendix \ref{app:robust} where we perform our rotation curve fits using a flat prior on $\eta$ over the range $0 < \eta < 2.75$ (the upper bound on the $\eta$ prior is set by energetic arguments; see Appendix \ref{app:robust} for details). This allows both no core ($\eta = 0$), corresponding to an NFW profile, and substantially larger cores than were found in the R16a simulations. In Appendix \ref{app:robust}, we show that the NFW profile ($\eta = 0$) is disfavoured at $>99$\% confidence, reaffirming the well-known cusp-core problem (see \S\ref{sec:intro}). However, as we showed for WLM in R16b, $\eta$ is otherwise poorly constrained (though consistent with our default choice of $\eta = 1.75$). Allowing $\eta$ to vary slightly increases our errors on $M_{200}$ but is otherwise benign. This is because $M_{200}$ is set by the outermost bins of the rotation curve where in many cases it begins to turn over and become flat. Indeed, in \S\ref{sec:exrots}, we show that demanding an NFW profile leads to a poor rotation curve fit, but little change in our derived halo masses. This demonstrates that so long as the dark matter density profile $\rho_{\rm dm}$ approaches the NFW form for $r > R_{1/2}$, our measurements of $M_{200}$ are not sensitive to our particular \coreNFW\ parameterisation of $\rho_{\rm dm}$.

\begin{table*}
\rotatebox{90}{
\resizebox{\textwidth}{!}{
\begin{minipage}{\textheight}
\hspace{-7mm}
\begin{tabular}{L{1.5cm} | c c c c c c c | c c c c | L{1.5cm} | l}
\hline
\hline
{\bf Galaxy} \vspace{1mm} & ${\mathbf v_{\rm max}}$ & ${\mathbf i}$ & $\mathbf{D}$ & $\mathbf{M_*}$ & $\mathbf{M_{\rm gas}}$ & $\mathbf{R_*}$ & $\mathbf{R_{\rm gas}}$ & $\mathbf{[R_{\rm min}, R_{\rm max}]}$ & $\mathbf{M_{200}}$ & $\mathbf{c}$ & $\mathbf{\chi^2_{\rm red}}$ & {\bf HI bubbles} & {\bf Refs.} \\
& (km/s) & ($^\circ$) & (Mpc) & $(10^7\,{\rm M}_\odot)$ & $(10^7\,{\rm M}_\odot)$ & (kpc) & (kpc) & (kpc) & $(10^{10} {\rm M}_\odot)$ &  &  &  & \\
\hline
\hline
\rowcolor{gray!40}NGC 6822 & $56.0 \pm 2.2$ & 65-75 & $0.49\pm 0.04$ & $7.63 \pm 1.9$ & $17.4$ & 0.68 & 1.94 & $[2.5,-]$ & $2^{+0.2}_{-0.3}$ & $15.1^{+1.8}_{-0.8}$ & 0.37 & $1.4 - 2$\,kpc; 0\,km/s & 1,2,3,7\\ [2ex]

WLM & $39.0 \pm 3.3$ & $74 \pm 2.3$ & $0.985\pm 0.033$ & $1.62 \pm 0.4$ & $7.9$ & $0.75$ & $1.04$ & $[0,-]$ & $0.83_{-0.2}^{0.2}$ & $17^{+3.9}_{-2.2}$ & 0.27 & 0.46\,kpc; 0\,km/s & 3,4,5,6,8 \\ [2ex]

\rowcolor{gray!40}DDO 52 & $50.7\pm13.4$ & $55.1 \pm 2.9$ & 10.3 & $5.27 \pm 1.3$ & $37.1$ & 0.94 & 2.49 & $[0,-]$ & $1.2_{-0.27}^{+0.29}$ & $17.3^{+4.2}_{-2.4}$ & 0.2 & -- & 3,6 \\ [2ex]

DDO 87 & $52.0\pm9.1$ & $42.7 \pm 7.3$ & 7.4 & $3.3 \pm 0.8$ & $31.0$ & 1.13 & 1.51 & $[0,-]$ & $1.13_{-0.25}^{+0.27}$ & $17.6_{-2.7}^{+4.6}$ & 0.31 & -- & 3,6 \\ [2ex]

\rowcolor{gray!40}DDO 154$^\bullet$ & $46.7\pm5.1$ & $67.9 \pm 1.1$ & 3.7 & $0.835 \pm 0.2$ & $30.9$ & 0.54 & 2.34 & $[0,-]$ & $1.26_{-0.05}^{+0.05}$ & $14.2_{-0.17}^{+0.24}$ & 2.14 & -- & 3,6 \\ [2ex]

DDO 210 & $17.8\pm9.5$ & $63.2 \pm 3.2$ & 0.9 & $0.068 \pm 0.017$ & $0.33$ & 0.22 & 0.25 & $[0.2,-]$ & $0.068_{-0.04}^{+0.13}$ & $21.4_{-5.2}^{+5.5}$ & 0.65 & -- & 3,6 \\ [2ex]

\rowcolor{gray!40}NGC 2366 & $58.8\pm5.4$ & $65.1 \pm 4.2$ & 3.4 & $6.95 \pm 1.73$ & $103$ & 1.54 & 2.69 & $[0,-]$ & $2.4_{-0.54}^{+0.49}$ & $17.3_{-2.4}^{+4.5}$ & 0.48 & $0.5-1$\,kpc; $30$\,km/s & 3,6,10 \\ [2ex]

UGC 8508 & $32.1\pm6.2$ & $67.6\pm5.3$ & 2.6 & $0.764 \pm 0.191$ & $3.5$ & 0.31 & 1.1 & $[0,-]$ & $0.63_{-0.2}^{+0.5}$ & $21.0_{-4.8}^{+5.3}$ & 0.05 & 0.285\,pc; 0\,km/s & 3,6,11 \\ [2ex]

\rowcolor{gray!40}CVnIdwA & $22.4 \pm 3.9$ & $49.2 \pm 10.9$ & 3.6 & $0.41 \pm 0.1$ & $6.42$ & 0.68 & 1.18 & $[0,-]$ & $0.17_{-0.05}^{+0.1}$ & $21.4_{-5.3}^{+5.4}$ & 0.21 & -- & 3,6 \\ [2ex]

DDO 126 & $39.2\pm3.1$ & $62.2 \pm 2.9$ & 4.9 & $1.61 \pm 0.4$ & $18.7$ & 0.82 & 1.51 & $[2,-]$ & $0.58_{-0.1}^{+0.2}$ & $20.6_{-4.7}^{+5.9}$ & 0.16 & -- & 3,6 \\ [2ex]

\rowcolor{gray!40}DDO 168 & $56.3 \pm 7.2$ & $47.3 \pm 7.4$ & 4.3 & $5.9 \pm 1.48$ & $45.8$ & 0.82 & 1.51 & $[0,-]$ & $2.1_{-0.48}^{+0.52}$ & $16.8_{-2.0}^{+3.2}$ & 0.24 & -- & 3,6 \\ [2ex]

\hline
\hline
\multicolumn{14}{c}{$\sim\sim\sim$ {\bf $i$-Rogues} $\sim \sim \sim$} \\
\hline
\hline

\rowcolor{gray!40}DDO 133 & $49.0\pm5.1$ & $36.9_{-4.0}^{+2.3}$ & 3.5 & $3.04 \pm 0.76$ & $16.9$ & 0.804 & 1.39 & $[0,-]$ & $1.6_{-0.44}^{+1.1}$ & $25.8_{-5.5}^{+2.8}$ & 0.16 & -- & 3,6 \\ [2ex]

IC 1613 & $19.9\pm2.2$ & $27.6_{-11.3}^{+9.0}$ & $0.74 \pm 0.01$ & $1.5 \pm 0.5$ & $8$ & 0.65 & 1.29 & $[1.9,-]$ & $0.17_{-0.1}^{+1.0}$ & $21.8_{-5.4}^{+5.3}$ & 0.13 & $1$\,kpc; 25\,km/s & 3,6,9 \\ [2ex]

\rowcolor{gray!40}DDO 50 & $37.6\pm10.1$ & $37.4_{-3.7}^{+2.0}$ & 3.4 & $10.72 \pm 2.68$ & $97.65$ & 0.89 & 3.1 & $[0,-]$ & $0.32_{-0.08}^{+0.16}$ & $26.2_{-4.7}^{+2.5}$ & 0.8 & 0.65-0.84\,kpc; 14\,km/s & 3,6,13 \\ [2ex]

DDO 53$^{\bullet}$ & $23.2\pm6.6$ & $23.9_{-6.4}^{+7.3}$ & 3.6 & $0.97 \pm 0.24$ & $24.6$ & 0.89 & 1.14 & $[0,-]$ & $0.39_{-0.36}^{+3.4}$ & $21.6_{-5.1}^{+5.3}$ & 0.38 & -- & 3,6 \\ [2ex]

\rowcolor{gray!40}DDO 47 & $62.6 \pm 5.2$ & $31.4_{-5.2}^{+5.7}$ & 5.2 & $9.4 \pm 4.7$ & $35.3$ & 0.7 & 8.2 & $[3.5,-]$ & $4.4_{-1.9}^{+3.2}$ & $20.5_{-4.6}^{+5.6}$ & 0.83 & -- & 6,23 \\ [2ex]

\hline
\hline
\multicolumn{14}{c}{$\sim\sim\sim$ {\bf Disequilibrium Rogues} $\sim \sim \sim$} \\
\hline
\hline

\rowcolor{gray!40}DDO 216 & $13.8\pm5.0$ & $70.0\pm5.0$ & 1.1 & $1.52 \pm 0.38$ & $0.152$ & 0.52 & 0.297 & $[0,-]$ & $0.067_{-0.025}^{+0.048}$ & $23.4_{-6.0}^{+4.6}$ & 0.52 & -- & 3,6 \\ [2ex]

NGC 1569 & $55.9 \pm 22.4$ & $67.0 \pm 5.6$ & 3.4 & $36 \pm 9$ & $23.8$ & 0.45 & 1.22 & $[1.8,-]$ & $0.81_{-0.7}^{+2.1}$ & $19.7_{-6.1}^{+6.7}$ & -- & ${\sim}200$\,pc; 10-70\,km/s & 3,6,14 \\ [2ex]

\hline
\hline
\multicolumn{14}{c}{$\sim\sim\sim$ {\bf Distance Rogues} $\sim \sim \sim$} \\
\hline
\hline

DDO 101 & $64.8\pm2.6$ & $52.4 \pm 1.7$ & 6.4 & $6.54 \pm 1$ & $3.48$ & 0.58 & 1.01 & $[0,-]$ & $5.2^{+0.6}_{-0.4}$ & $28.9_{-1.3}^{+0.6}$ & 7.2 & -- & 3,6 \\ [1ex]
& & & 12.9 & $26.6 \pm 4$ & $14.13$ & 1.16 & 2.03 & $[0,-]$ & $3.0^{+0.4}_{-0.2}$ & $28.3_{-2.2}^{+1.1}$ & 1.92 & -- & -- \\ [2ex]

\hline
\hline
\multicolumn{14}{c}{$\sim\sim\sim$ {\bf No Rotation Curve} $\sim \sim \sim$} \\
\hline
\hline

Leo T & -- & ${\sim}20^\circ$ & $0.407 \pm 0.038$ & $0.0135 \pm 0.008$ & $0.028$ & 0.106 & -- & -- & $0.035-0.075$ & -- & -- & -- & 16,17,18,19,20 \\  [2ex]

\rowcolor{gray!40}Carina & -- & -- & $0.085 \pm 0.005$ & $0.048_{-0.004}^{+0.006}$ & -- & 0.177 & -- & -- & $0.036_{-0.023}^{+0.038}$ & -- & -- & -- & 15,21,22 \\  [2ex]

\hline
\hline
\end{tabular}
\end{minipage}
}
}
\vspace{2mm}
\caption{Data and derived rotation curve fitting parameters for 19 isolated dIrrs and two Milky Way satellite dwarfs. The dIrrs are divided into a clean sample, inclination `rogues', disequilibrium rogues and distance rogues, as marked (see \S\ref{sec:rogues} for more details). We highlight galaxies with high gas fractions $M_{\rm gas} / M_* > 20$ with a `$\bullet$'. Column 1 gives the galaxy name. Columns 2-7 give the data for that galaxy: the peak asymmetric drift corrected rotation curve velocity $v_{\rm max}$; the distance to the galaxy $D$; the stellar mass, with errors $M_*$; the total gas mass $M_{\rm gas}$; and the exponential stellar and gas disc scale lengths $R_*$ and $R_{\rm g}$, respectively. Column 8 gives the radial range used in the fit to the rotation curve $[R_{\rm min}, R_{\rm max}]$ (`$-$' indicates that $R_{\rm max}$ is set to the outermost data point). Columns 9-10 give the marginalised dark matter halo parameters: the virial mass $M_{200}$ and concentration parameter $c$, with 68\% confidence intervals. Column 11 gives the reduced $\chi^2_{\rm red}$ of the fit. Column 12 reports data on the presence of HI bubbles that can indicate disequilibria (see R16b). The size and expansion velocity of the largest bubble are given (``--" denotes no reported measurement). Finally, column 12 gives the data references for that galaxy, as follows: 1: \citet{1884AN....110..125B}; 2: \citet{2003MNRAS.340...12W}; 3: \citet{2012AJ....143...47Z}; 4: \citet{2011AJ....141..194G}; 5: \citet{2012ApJ...750...33L}; 6: \citet{2015AJ....149..180O}; 7: \citet{2000ApJ...537L..95D}; 8: \citet{2007AJ....133.2242K}; 9: \citet{2006AA...448..123S}; 10: \citet{2009AA...493..511V}; 11: \citet{2011ApJ...738...10W}; 12: \citet{2013AJ....146...42A}; 13: \citet{1992AJ....103.1841P}; 14: \citet{2012AJ....144..152J}; 15: \citet{2015NatCo...6E7599U}; 16: \citet{2008ApJ...684.1075M}; 17: \citet{2012ApJ...748...88W}; 18: \citet{2008MNRAS.384..535R}; 19: \citet{2013ApJ...779..102K}; 20: \citet{2015arXiv150804143R}; 21: \citet{1995MNRAS.277.1354I}; 22: \citet{2014A&A...572A..10D}; 23: \citet{1999A&AS..139..491M}. The last two rows show data for the isolated dwarf irregular Leo T and the Milky Way dwarf spheroidal galaxy Carina, neither of which have gaseous rotation curves. We estimate $M_{200}$ for Leo T by direct comparison to the simulations in R16a (see \S\ref{sec:data}); for Carina, we use the pre-infall `tidal mass estimate' from \citet{2015NatCo...6E7599U}.}
\label{tab:data}
\end{table*}

\subsection{Fitting the mass model to data \& our choice of priors}\label{sec:mcmc}

We fit the above mass model to the data using the \EMCEE\ affine invariant Markov Chain Monte Carlo (MCMC) sampler from \citet{2013PASP..125..306F}. We assume uncorrelated Gaussian errors such that the Likelihood function is given by $\mathcal{L} = \exp(-\chi^2/2)$. We use 100 walkers, each generating 1500 models and we throw out the first half of these as a conservative `burn in' criteria. We explicitly checked that our results are converged by running more models and examining walker convergence. All parameters were held fixed except for the dark matter virial mass $M_{200}$; the concentration parameter $c$; and the total stellar mass $M_*$. We assume a flat logarithmic prior on $M_{200}$ of $8 <  \log_{10}\left[M_{200}/{\rm M}_\odot\right] < 11$; a flat linear prior on $c$ of $14 < c < 30$ and a flat linear prior on $M_*$ over the range given by stellar population synthesis modelling, as reported in Table \ref{tab:data}. We assume an error on $M_*$ of $25\%$ unless a larger error than this is reported in the literature \citep{2012AJ....143...47Z,2015AJ....149..180O}. The generous prior range on $c$ is set by the cosmic mean redshift $z=0$ expectation value of $c$ at the extremities of the prior on $M_{200}$ \citep{2007MNRAS.378...55M}. In R16b, we showed that our results are not sensitive to this prior choice. For each galaxy, we fit data over a range $[R_{\rm min}, R_{\rm max}]$ as reported in Table \ref{tab:data}, where `$-$' means that $R_{\rm max}$ was set by the outermost data point. $R_{\rm min}$ is marked by thin vertical lines on the individual rotation curve fits reported in Appendix \ref{app:rotcurves}. For most galaxies, $R_{\rm min} = 0$. It is only non-zero where the innermost rotation curve is affected by an expanding HI bubble (see R16b for further details). In Appendix \ref{app:robust}, we explore allowing the core size parameter $\eta$ (equation \ref{eqn:etarc}) to vary also in the fits.

\subsection{Tests on mock data and the exclusion of `rogues'}\label{sec:rogues}

Our ability to measure $M_{200}$ and $c$ from mock rotation curve data was extensively tested in R16b. There, we showed that there are three key difficulties that can lead to systematic biases. Firstly, we must account for cusp-core transformations due to stellar feedback if we wish to obtain a good fit to the rotation curve inside $R_{1/2}$. We account for this by using our \coreNFW\ profile (\S\ref{sec:massmodel}). Secondly, our simulated dwarfs continuously cycle between quiescent and `starburst' modes that cause the HI rotation curve to fluctuate. This can lead to a systematic bias on $M_{200}$ of up to half a dex in the most extreme cases. However, this disequilibrium can be readily identified by the presence of large and fast-expanding (${\simgt}20-30$\,km\,s$^{-1}$) HI superbubbles in the ISM. Thirdly, low inclination galaxies, particularly if also undergoing a starburst, can be difficult to properly inclination correct. Using mock HI datacubes, we found that \Barolo\ can return a systematically low inclination if $i_{\rm fit} \simlt 40^\circ$. For this reason, if \Barolo\ returns an inclination of $i_{\rm fit} < 40^\circ$, we marginalise over $i$ in our fits assuming a flat prior over the range $0^\circ < i < 40^\circ$. We call such galaxies `inclination Rogues' or $i$-Rogues and we discuss them in Appendix \ref{app:robust}. (We find that five of our 19 dIrrs are $i$-Rogues.)

Two galaxies -- DDO 216 (Pegasus) and NGC 1569 -- have highly irregular rotation curves. For Pegasus, this owes to the limited radial extent of its rotation curve that does not extend beyond $R_{1/2}$. For NGC 1569, its inner rotation curve is shallower than required to support even its stellar mass, indicating that it is far from equilibrium. This is further supported by the presence of large and fast-expanding HI holes (see R16b; Table \ref{tab:data}; and \citealt{2012AJ....144..152J}) and the fact that it is classified as a `Blue Compact Dwarf', with a very recent starburst some ${\sim}40$\,Myrs ago \citep{2010ApJ...721..297M}. (Indeed, \citet{2014A&A...566A..71L} classify it as having a `kinematically disturbed HI disc' and do not attempt to derive its rotation curve.) From a more theoretical standpoint, R16b and more recently \citet{2016arXiv161004232E} show that starbursts are expected to drive exactly the sort of disequilibrium seen in NGC 1569. For these reasons, we exclude these two `disequilibrium rogues' from further analysis from here on. For completeness, we report their best-fitting $M_{200}$ and $c$ in Table \ref{tab:data} and we show their rotation curve fits in Appendix \ref{app:rotcurves}.

Finally, one galaxy -- DDO 101 -- has a very uncertain distance; we refer to this galaxy as a `distance Rogue'. We discussed DDO 101 in detail in R16b, showing that for a distance of ${\sim}12$\,Mpc it can be well-fit by a \coreNFW\ dark matter halo. We consider its position on the \MstarMrot\ relation alongside the $i$-Rogues in Appendix \ref{app:robust}.

\section{Results}\label{sec:results}

\subsection{The rotation curve fits}\label{sec:exrots}

\begin{figure*}
\begin{center}
\includegraphics[width=0.99\textwidth]{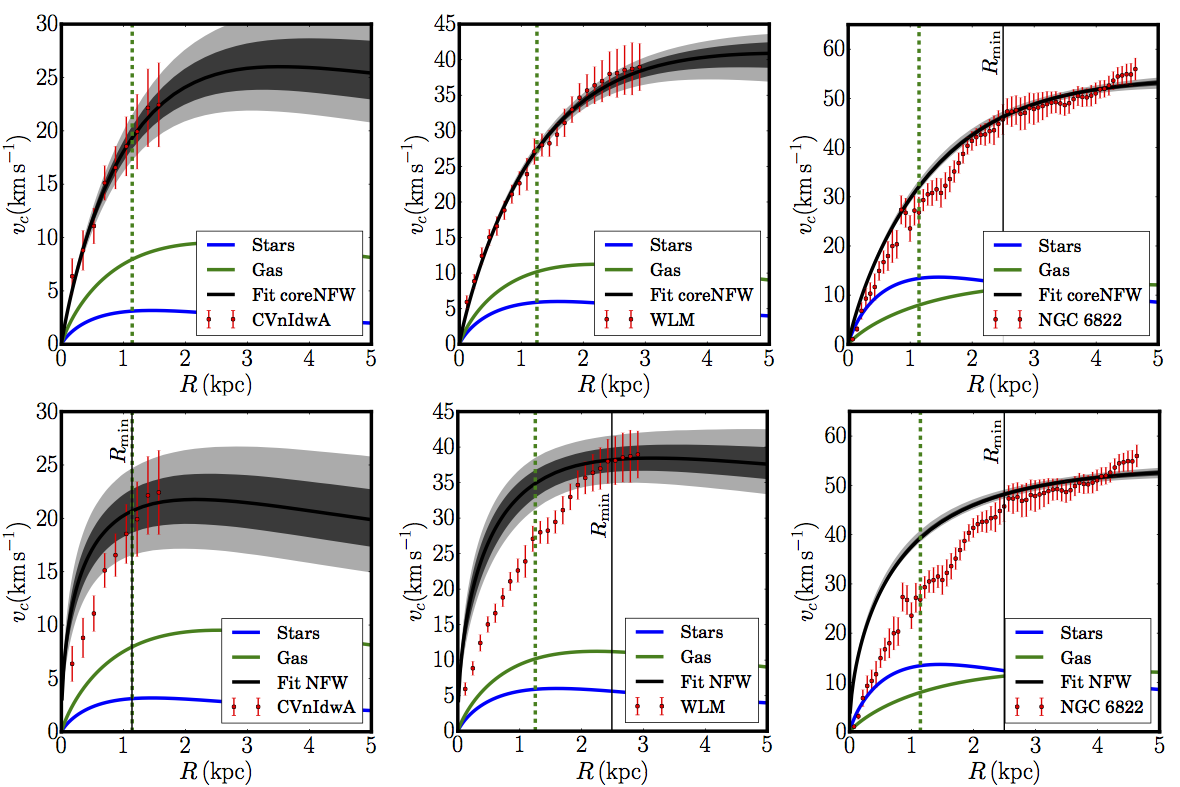}
\caption{Rotation curve fits for three example galaxies: CVnIdwA, WLM and NGC 6822, chosen to span the range of stellar masses in our full sample (see Table \ref{tab:data}). We show the full sample, including the `rogues', in Appendix \ref{app:rotcurves}. The black contours show the median (black), 68\% (dark grey) and 95\% (light grey) confidence intervals of our fitted rotation curve models (see \S\ref{sec:massmodel}). The vertical green dashed line shows the projected stellar half light radius $R_{1/2}$. The thin vertical black line marks the inner data point used for the fit, $R_{\rm min}$ (where this is not marked $R_{\rm min} = 0$). The blue and green lines show the rotation curve contribution from stars and gas, respectively. The top three panels show fits using our \coreNFW\ profile that accounts for cusp-core transformations due to stellar feedback (see \S\ref{sec:massmodel}). These give an excellent fit to the rotation curve shape in all three cases. The bottom three panels show fits using an \NFW\ profile, where we set $R_{\rm min}$ to ensure that the outer rotation curve is well-fit. This gives a much poorer fit to the rotation curve shape, reaffirming the long-standing `cusp-core' problem.}
\label{fig:exrots} 
\end{center}
\end{figure*}

In Figure \ref{fig:exrots}, we show three example rotation curve fits for CVnIdwA, WLM and NGC 6822, chosen to span the range of stellar masses in our full sample (see Table \ref{tab:data}). (We show the full sample, including the `rogues', in Appendix \ref{app:rotcurves}.) The black contours show the median (black), 68\% (dark grey) and 95\% (light grey) confidence intervals of our fitted rotation curve models (see \S\ref{sec:massmodel}). The vertical green dashed line shows the projected stellar half light radius $R_{1/2}$. The thin vertical black line marks the inner data point used for the fit, $R_{\rm min}$ (where this is not marked $R_{\rm min} = 0$). The blue and green lines show the rotation curve contribution from stars and gas, respectively. The top three panels of Figure \ref{fig:exrots} show fits using the \coreNFW\ profile, the bottom three using an \NFW\ profile, where we set $R_{\rm min}$ to ensure that the outer rotation curve is well-fit (see \S\ref{sec:rotmethod} for details of our fitting methodology and priors). 

As can be seen in Figure \ref{fig:exrots}, in all three cases the \coreNFW\ profile provides an excellent fit to the data, while the \NFW\ profile gives a poor fit, reaffirming the long-standing `cusp-core' problem (see \S\ref{sec:intro}). The good fits that we find when using the \coreNFW\ profile are particularly striking since, like the \NFW\ profile, it has only two free parameters: $M_{200}$ and $c$ (see \S\ref{sec:massmodel}). However, despite the \NFW\ profile giving a poor fit to the rotation curve shape, the NFW-derived $M_{200}$ are actually in good agreement with those from our \coreNFW\ fits. For NGC 6822, we find $M_{200,{\rm NFW}} = 2.0_{-0.2}^{+0.13} \times 10^{10}$\,M$_\odot$; for WLM, $M_{200,{\rm NFW}} = 5.2_{-1.2}^{+2.1} \times 10^{9}$\,M$_\odot$; and for CVnIdwA, $M_{200,{\rm NFW}} = 0.79_{-0.3}^{+0.5} \times 10^{9}$\,M$_\odot$. These agree, within our 68\% confidence intervals, with the \coreNFW\ values reported in Table \ref{tab:data}.

The above demonstrates that the \coreNFW\ profile is important for obtaining a good fit to the rotation curve shape inside ${\sim}R_{1/2}$, however it is not critical for measuring $M_{200}$. What matters for measuring $M_{200}$ is that the dark matter density profile approaches the \NFW\ form for $r > R_{1/2}$ (as is the case for the \coreNFW\ profile by construction). Since there is not enough integrated supernova energy to unbind the dark matter cusp on scales substantially larger than $R_{1/2}$ (see e.g. \citealt{2012ApJ...759L..42P}; R16a; and \S\ref{sec:discussion}), this demonstrates that our results for $M_{200}$ are robust to the details of stellar feedback-induced dark matter heating. We confirm this in Appendix \ref{app:robust}, where we show that allowing the dark matter core size to vary freely in the fits slightly inflates the errors on $M_{200}$, but does not otherwise affect our results.

\subsection{The stellar mass-halo mass relation of isolated field dwarfs}\label{sec:mstarmrot}

\begin{figure}
\begin{center}
\includegraphics[width=0.49\textwidth]{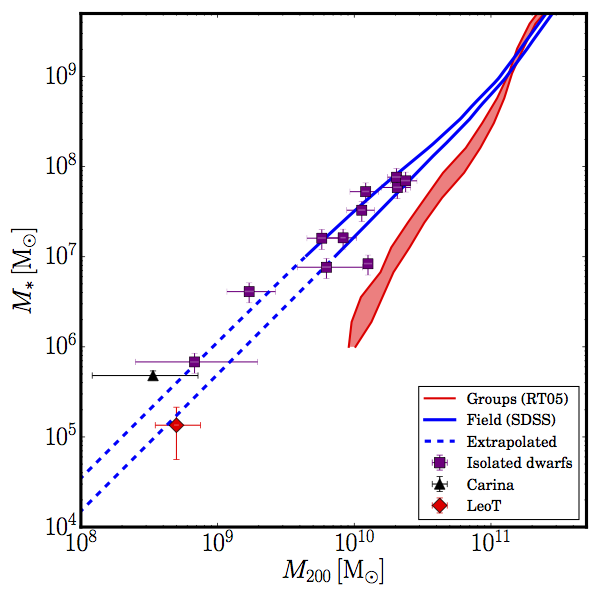}
\caption{The stellar mass-halo mass relation of 11 isolated dIrr galaxies, derived from their HI rotation curves (\MstarMrot); and two galaxies that do not have HI rotation curves: the isolated dwarf irregular Leo T (diamond) and the Milky Way dwarf spheroidal galaxy Carina (black triangle). (The masses of these two galaxies are derived as described in \S\ref{sec:data}.) All data are reported in Table \ref{tab:data}. Overplotted are \MstarMabund\ calculated from abundance matching in $\Lambda$CDM using the SDSS field stellar mass function (solid blue lines) and the RT05 stellar mass function of nearby groups (red shaded region). The lines are dashed where they rely on a power law extrapolation of the SDSS stellar mass function below $M_* \sim 10^7$\,M$_\odot$. Notice that \MstarMabund\ (blue lines) gives a remarkable match to \MstarMrot\ (purple data points) down to $M_{200} \sim 5 \times 10^9$\,M$_\odot$, and $M_{200} \sim 5 \times 10^8$\,M$_\odot$ if we use the power law extrapolation of the SDSS stellar mass function (dashed lines). However, \MstarMabund\ derived from the stellar mass function of nearby galaxy groups (red shaded region) gives a poor match.}
\label{fig:carina_plot}
\end{center}
\end{figure}

In Figure \ref{fig:carina_plot}, we plot the stellar mass-halo mass relation of the 11 `clean' isolated dIrrs listed in Table \ref{tab:data}, as derived from their HI rotation curves (see \S\ref{sec:rotmethod}). For our `clean' sample, we include all galaxies with inclination $i_{\rm fit} > 40^\circ$; well measured distance; and no obvious signs of disequilibrium (see \S\ref{sec:rogues}). The individual rotation curves for these galaxies are reported in Appendix \ref{app:rotcurves}. There, we also show the rotation curves for the `rogues' that did not make the above cut (see Table \ref{tab:data}). In addition, on Figure \ref{fig:carina_plot} we plot two galaxies that do not have HI rotation curves: the isolated dwarf irregular Leo T (red diamond) and the Milky Way dwarf spheroidal galaxy Carina (black triangle). We estimate $M_{200}$ for Leo T by direct comparison to the simulations in R16a (see \S\ref{sec:data} and Table \ref{tab:data}); for Carina, we use the pre-infall `tidal mass estimate' from \citet{2015NatCo...6E7599U}. We discuss Carina further in \S\ref{sec:carina}. Overplotted on Figure \ref{fig:carina_plot} are \MstarMabund\ calculated from abundance matching in $\Lambda$CDM using the SDSS field stellar mass function (solid blue lines) and the RT05 stellar mass function of nearby  groups (red shaded region). We discuss these in \S\ref{sec:mstarmabund}. 

Notice from Figure \ref{fig:carina_plot} that the isolated dwarfs show remarkably little scatter, defining a monotonic line in \MstarMrot\ space within their 68\% confidence intervals. Such monotonicity is a key assumption of abundance matching and Figure \ref{fig:carina_plot} demonstrates that this assumption is empirically justified, at least for the sample of isolated dIrrs that we consider here. There is, however, one significant outlier, DDO 154. We discuss this interesting galaxy further in \S\ref{sec:discussion}.

\subsection{The stellar mass function in groups and in the field}\label{sec:stellarmfuncs}

In this section, we compare four stellar mass functions taken from the literature, as reported in Figure \ref{fig:stellar_massfunc_compare}. The blue data points show the stellar mass function from SDSS \citep{2013ApJ...770...57B}; the red data points from GAMA \citep{2012MNRAS.421..621B}; the black data points from RT05; and the magenta data points from zCOSMOS \citep{2012A&A...538A.104G}. (See \S\ref{sec:data} for a description of these surveys.) The green tracks show a non-parametric fit to the SDSS stellar mass function, where the upper and lower tracks encompass the 68\% confidence intervals of the data. Below $M_* = 10^8$\,M$_\odot$, we fit a single power law to the SDSS data and use this to extrapolate to lower stellar mass. As in \citet{2008MNRAS.388..945B}, we find a best-fit logarithmic slope of $\alpha = 1.6$, where $\left.dN/dM\right|_{M_* < 10^8\,{\rm M}_\odot} \propto M^{-\alpha}$.

\subsubsection{Evidence for a shallower group stellar mass function below $M_* \sim 10^9$\,M$_\odot$}

Firstly, notice that below $M_* \sim 10^9$\,M$_\odot$ the SDSS and RT05 stellar mass functions diverge, with the RT05 mass function becoming substantially shallower. This difference has been noted previously in the literature \citep[e.g.][]{2008MNRAS.388..945B} but to date has remained unexplained. Here, we suggest that it owes to an {\it environmental} dependence. The RT05 stellar mass function was built using a compilation of SDSS data at the bright end, and the luminosity function of the \citet{2002MNRAS.335..712T} {\it local groups catalogue} at the faint end (see \S\ref{sec:stellarmassfunc}). Thus, by construction, below $M_* \sim 10^9$\,M$_\odot$ RT05 measured the stellar mass function of nearby galaxy groups. Indeed, we find further evidence for this from the zCOSMOS survey. \citet{2012A&A...538A.104G} split the zCOSMOS stellar mass function into a `field galaxy' sample (Figure \ref{fig:stellar_massfunc_compare}; magenta circles) and an X-ray selected group sample (magenta diamonds) over the redshift range $z=0.2-0.4$. As can be seen in Figure \ref{fig:stellar_massfunc_compare}, the zCOSMOS stellar mass functions are only complete down to $M_* \sim 4 \times 10^8$\,M$_\odot$ (see \S\ref{sec:data}) but nonetheless, at this mass scale, there is a statistically significant bifurcation between the zCOSMOS field and group samples that matches that seen in SDSS and RT05.

There are two key challenges involved in comparing zCOSMOS with RT05 and SDSS. Firstly, the redshift range of the surveys are different (see \S\ref{sec:data}). However, this is not a significant effect since the stellar mass function is known to be almost constant out to $z = 0.5$ \citep{2013ApJ...770...57B}. (Indeed, we find no difference between the zCOSMOS field stellar mass function (magenta circles) and that of SDSS (blue data points) down to $M_* \sim 4 \times 10^8$\,M$_\odot$.) Secondly, the definition of a `group' differs. The \citet{2012A&A...538A.104G} groups are selected based on co-added XMM and Chandra X-ray images, using a wavelet method to detect extended emission \citep{2007ApJS..172..182F}. In this way, they find groups over the mass range $0.14 < M_{500}/(10^{13}{\rm M}_\odot) < 26$. By contrast, \citet{2002MNRAS.335..712T} study five nearby optically selected groups, including the Local Group. Only one of these has reported X-ray emission \citep{2009AJ....137.4956R}, but they do span a similar mass range to the \citet{2012A&A...538A.104G} sample \citep{2007ApJ...656..805Z,2011MNRAS.412.2498M,2016MNRAS.456L..54P}. For these reasons, a direct comparison between the \citet{2002MNRAS.335..712T} and \citep{2012A&A...538A.104G} groups is reasonable. Indeed, their stellar mass functions agree remarkably well down to the stellar mass limit of the zCOSMOS survey (compare the magenta diamonds and black data points in Figure \ref{fig:stellar_massfunc_compare}).

Finally, consider the GAMA stellar mass function in Figure \ref{fig:stellar_massfunc_compare} (red data points). This agrees well with both the zCOSMOS field sample (magenta circles) and SDSS (blue data points) down to $M_* \sim 2 \times 10^8$\,M$_\odot$. The one data point below this is slightly, though not statistically significantly, shallower than SDSS. It is beyond the scope of this present work to explore this discrepancy in any detail, though it has been noted previously \citep[e.g.][]{2012MNRAS.421..621B}. As emphasised in \S\ref{sec:data}, the GAMA survey covers about one tenth of the volume of SDSS and is complete only at a higher stellar mass. For these reasons, we will use the SDSS stellar mass function for the remainder of this paper. We discuss the GAMA stellar mass function further in \S\ref{sec:discussion}.

\subsubsection{The origin of the $M_* \sim 10^9\,$M$_\odot$ mass scale}\label{sec:mquench}

The shallower group stellar mass function that we find here is perhaps not surprising. It has long been known that satellites are quenched on infall to groups due to a combination of ram pressure stripping and tides \citep[e.g.][]{2012ApJ...757....4P,2013MNRAS.433.2749G,2013ApJ...776...71C}. Ram pressure shuts down star formation, leading to a lower stellar mass for a given pre-infall halo mass, while tides physically destroy halos depleting the dark matter subhalo mass function. In the Milky Way, this is evidenced by the `distance-morphology' relation: most satellites within ${\sim}200$\,kpc of the Galactic centre have truncated star formation and are devoid of gas, while those at larger radii have HI and are currently forming stars \citep[e.g.][]{1984ARA&A..22..445H,1998ARA&A..36..435M,2001ApJ...559..754M,2003AJ....125.1926G,2009ARA&A..47..371T,2012AJ....144....4M,2013MNRAS.433.2749G}.

It is interesting to ask, however, whether ram pressure or tides can explain why the stellar mass function is affected only below $M_* \sim 10^9$\,M$_\odot$. In R16a, we calculated the effect of tides on satellites orbiting within a Milky Way mass host (their section 4.3). The effect is maximised if satellites have their dark matter cusps transformed into cores. But even in this extremum limit, satellites are only fully destroyed if they have a pericentre of $r_p \simlt 30$\,kpc and a mass $M_{200} \simlt 10^{10}$\,M$_\odot$. Using our \MstarMrot\ relation in Figure \ref{fig:carina_plot}, this corresponds to a stellar mass of $M_* \sim 2-3 \times 10^7$\,M$_\odot$, suggesting that tides are not likely to be the primary cause of the shallower group stellar mass function that we find here. 

The second potential culprit is ram pressure. This occurs when \citep{2013MNRAS.433.2749G}: 

\begin{equation} 
\rho_h(r_p) v_p^2 \simgt \frac{1}{5}\rho_d \frac{v_{\rm max}^2}{2}
\label{eqn:rampressure}
\end{equation}
where $\rho_h$ is the coronal gas density of the host at pericentre; $v_p$ is the velocity of the satellite at pericentre; $\rho_d$ is the density of gas in the dwarf ISM; $v_{\rm max}$ is the peak rotational velocity of the dwarf\footnote{We have assumed here that $v_{\rm max}^2 \simeq 2 \sigma_*^2$, where $\sigma_*$ is the stellar velocity dispersion of the dwarf. This amounts to an assumption of a flat, isothermal, rotation curve for the dwarf \citep[e.g.][]{1987gady.book.....B}.}; and the factor $1/5$ accounts for non-linear effects \citep{2013MNRAS.433.2749G}.  

For a satellite falling into the Milky Way, $\rho_h \sim 3 \times 10^{-4}$\,atoms\,cm$^{-3}$; $v_p \sim 450$\,km/s; and $\rho_d \sim 0.1$\,atoms\,cm$^{-3}$ \citep{2013MNRAS.433.2749G}. Thus, we can rearrange equation \ref{eqn:rampressure} to provide a limiting $v_{\rm max}$ below which ram pressure becomes important:

\begin{equation}
v_{\rm max,ram} = \sqrt{\frac{10\rho_h}{\rho_d}} v_p \sim 78\,{\rm km\,s}^{-1}
\end{equation} 
This is similar to the $v_{\rm max}$ of the LMC \citep{2002AJ....124.2639V} that has a stellar mass of $M_* \sim 1.5 \times 10^9$\,M$_\odot$ \citep{2012AJ....144....4M}. This suggests that the shallower group stellar mass function that we find here owes to satellite quenching, driven primarily by ram pressure. Indeed, \citet{2012ApJ...757...85G} found, using SDSS data, that {\it all} field galaxies above $M_* = 10^9$\,M$_\odot$ are star-forming today, independent of environment. By contrast, galaxies with $M_* < 10^9$\,M$_\odot$ can be quenched, with the quenched fraction increasing with proximity to a larger host galaxy.

\subsection{Abundance matching in groups and in the field}\label{sec:mstarmabund}

In this section, we measure \MstarMabund\ using the stellar mass functions in Figure \ref{fig:stellar_massfunc_compare} matched to the $\Lambda$CDM Bolshoi simulation \citep{2011ApJ...740..102K}. Our abundance matching is `non-parametric' in the sense that we numerically integrate the curves in Figure \ref{fig:stellar_massfunc_compare} to obtain the cumulative stellar mass function; we then match these numerically to the cumulative halo mass function from the Bolshoi simulation. For this latter, we use a Schechter function fit to the halo mass function, defining the `halo mass' as the virial mass $M_{200}$ before infall. 

The results are shown in Figure \ref{fig:carina_plot} for the SDSS field stellar mass function (blue lines) and the RT05 group stellar mass function (red shaded region). The lines are dashed where they rely on a power law extrapolation of the SDSS stellar mass function below $M_* \sim 10^7$\,M$_\odot$. (We compare and contrast our abundance matching results with previous determinations in the literature in Appendix \ref{app:abundcompare}.)

Notice that \MstarMabund\ (blue lines) gives a remarkable match to \MstarMrot\ (purple data points) down to $M_{200} \sim 5 \times 10^9$\,M$_\odot$, and $M_{200} \sim 5 \times 10^8$\,M$_\odot$ if we use the power law extrapolation of SDSS. However, \MstarMabund\ derived from the stellar mass function of nearby galaxy groups (RT05; red shaded region) gives a poor match. In particular, it leads to the familiar result that all dwarf galaxies must inhabit implausibly massive ${\sim}10^{10}$\,M$_\odot$ halos \citep[e.g.][]{2006MNRAS.tmp..153R} that has become known as the `too big to fail' (TBTF) problem \citep[e.g.][]{2011MNRAS.415L..40B}. 

There are a number of problems with using the RT05 stellar mass function for `classical' abundance matching as we have done here. Firstly, we have assumed a monotonic relation between $M_*$ and $M_{200}$. We have shown that this is true for our sample of isolated dIrrs (\S\ref{sec:mstarmrot}), but we expect it to fail for satellites whose $M_*$ will depend on $M_{200}$, their time of infall and their orbit, inducing scatter in $M_*$ for a given pre-infall $M_{200}$ \citep[e.g.][and see \S\ref{sec:mquench}]{2015NatCo...6E7599U,2016arXiv160500004T,2016arXiv160304855G}. Secondly, there is what we might call a `volume problem'. If we wish to match a pure-dark matter simulation to the Milky Way, what volume should we use to normalise the Milky Way satellite mass function? \citet{2014ApJ...784L..14B} and \citet{2014MNRAS.444..222G} solve this by explicitly matching satellites to constrained simulations of the Local Group. Here, we solve it by using the RT05 stellar mass function. This solves the `volume problem' by renormalising the group stellar mass functions derived from \citet{2002MNRAS.335..712T} to match SDSS at the bright end (see \S\ref{sec:data}). Since this normalises the volume to SDSS field galaxies, we must then abundance match RT05 with the full Bolshoi simulation, as we have done here. Indeed, in Appendix \ref{app:abundcompare} we verify that our RT05 \MstarMabund\ relation, derived in this way, agrees very well with those derived independently by \citealt{2014ApJ...784L..14B} and \citet{2014MNRAS.444..222G}. Finally, there is the problem of satellite quenching. As discussed in \S\ref{sec:mquench}, satellites can have their star formation shut down by ram pressure stripping, or be tidally disrupted on infall. Tidal stripping is already dealt with, in part, by using the pre-infall $M_{200}$ \citep[e.g.][]{2007ApJ...667..859D}. However, we expect tidal disruption to be enhanced  by cusp-core transformations and the presence of the Milky Way stellar disc, neither of which are captured by pure dark matter simulations \citep[e.g.][and see the discussion in R16a]{2006MNRAS.tmp..153R,2010MNRAS.406.1290P,2010ApJ...709.1138D,2012ApJ...761...71Z,2016arXiv160205957W}. 

For all of the above reasons, we expect `classical' abundance matching with the RT05 stellar mass function to fail. Nonetheless, it is instructive because the key assumptions that go into it are common in the literature \citep[e.g.][]{2014ApJ...784L..14B,2014MNRAS.444..222G}. Indeed, it is likely that these assumptions are responsible for the now long-standing `missing satellites' and TBTF problems that manifest for satellite galaxies below $M_{\rm TBTF} \sim 10^{10}$\,M$_\odot$ \citep[][and see \S\ref{sec:intro}]{2006MNRAS.tmp..153R,2011MNRAS.415L..40B,2014MNRAS.440.3511T}. The fact that \MstarMrot\ matches \MstarMabund\ for our sample of isolated dIrrs demonstrates that every isolated field halo is occupied with a dIrr down to $M_{200} \sim 5 \times 10^9$\,M$_\odot$ and to $M_{200} \sim 5 \times 10^8$\,M$_\odot$ if we use the power law extrapolation of SDSS. Furthermore, these dwarfs inhabit dark matter halos that are perfectly consistent with their observed gaseous rotation curves. Thus, our sample of isolated dIrrs -- that extend to $M_{200} < M_{\rm TBTF}$ -- has no missing satellites or TBTF problem, suggesting that both depend on {\it environment}. We discuss this further in \S\ref{sec:discussion}.

\subsection{Constraints on warm dark matter}\label{sec:wdm}

\begin{figure}
\begin{center}
\includegraphics[width=0.49\textwidth]{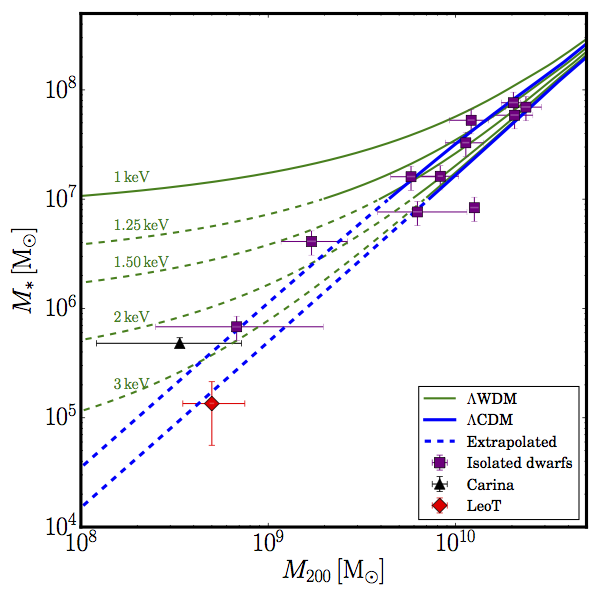}
\caption{\MstarMrot\ (purple data points) as compared to \MstarMabund\ in $\Lambda$CDM (blue lines) and $\Lambda$WDM (green lines), using the SDSS field stellar mass function. The thermal relic mass $m_{\rm WDM}$ is marked on the curves in keV. The lines and symbols are as in Figure \ref{fig:carina_plot}.}
\label{fig:carina_plot_wdm}
\end{center}
\end{figure}

We have shown so far that the field dIrr \MstarMabund\ is consistent with \MstarMrot\ in $\Lambda$CDM. In this section, we consider how well these match in a $\Lambda$WDM cosmology. We use the formulae in \citet{2012MNRAS.424..684S} to transform the Bolshoi halo mass function, derived for $\Lambda$CDM, to one in $\Lambda$WDM\footnote{
There is a known problem in the literature with the formation of spurious halos at the resolution limit in WDM simulations \citep[e.g.][]{2007astro.ph..2575W,2013MNRAS.tmp.1799H,2013MNRAS.434.3337A,2014MNRAS.439..300L,2015arXiv150302689H}. Equation \ref{eqn:wdmschneider} is derived from fits to $N$-body simulations where such spurious halos have been pruned from the analysis. We refer to it as describing an `effective warm dark matter' cosmology because it really describes a suppression in the halo mass function at low mass, parameterised by an effective thermal relic mass, $m_{\rm WDM}$. More realistic warm dark matter models will show model-specific features in the small scale matter power spectrum (see e.g. \citealt{2009JCAP...05..012B} for sterile neutrino models). It is beyond the scope of this present work to test such models in detail.
}:

\begin{equation} 
\left. \frac{d N}{dM}\right|_{\rm WDM} = \left. \frac{d N}{dM}\right|_{\rm CDM} \left(1 + \frac{M_{\rm hm}}{M}\right)^{-\beta}
\label{eqn:wdmschneider}
\end{equation}
where $M_{\rm hm} = 4/3\pi \rho_{\rm crit} (\lambda_{\rm hm}/2)^3$ is the `half mode mass'; $\beta = 1.16$; $\lambda_{\rm hm}$ is the `half mode' scale length, given by: 

\begin{equation} 
\lambda_{\rm hm} = 0.683 \left(\frac{m_{\rm WDM}}{{\rm keV}}\right)^{-1.11} \left(\frac{\Omega_{\rm M}}{0.25}\right)^{0.11}\left(\frac{h}{0.7}\right)^{1.22}\,{\rm Mpc}\,h^{-1}; 
\end{equation}
$m_{\rm WDM}$ is the warm dark matter particle mass in keV; $\Omega_{\rm M}$ is the matter density of the Universe at redshift $z=0$; and $h$ is the Hubble parameter (we assume the same cosmological parameters as in the Bolshoi simulation; see Table \ref{tab:cosmopars}).

In Figure \ref{fig:carina_plot_wdm}, we show tracks of \MstarMabund\ in $\Lambda$WDM for varying thermal relic mass over the range $1 < m_{\rm WDM} < 5$\,keV, as marked (green lines). Where these lines rely on the extrapolated SDSS stellar mass function, they are dashed. We deliberately pick the most conservative limits by using the lower bound of the SDSS stellar mass function to calculate \MstarMabund.

As can be seen from Figure \ref{fig:carina_plot_wdm}, without using the power law extrapolation of the SDSS stellar mass function below $M_* \sim 10^7$\,M$_\odot$, we can rule out $m_{\rm WDM} < 1.25$\,keV at 68\% confidence. Using the power law extrapolation, this limit improves to $m_{\rm WDM} < 2$\,keV at 68\% confidence. If we further add the Leo T data point, then this tightens to $m_{\rm WDM} < 3$\,keV. However, for this limit to become robust we would need to find many more Leo T-like galaxies in the Local Volume, ideally with measured rotation curves. We discuss this further in \S\ref{sec:discussion}.

Our limit on $m_{\rm WDM}$ approaches the latest limits from the Lyman-$\alpha$ forest \citep[e.g.][]{2015arXiv151201981B}. It is competitive with a more model-dependent limit from Local Group satellite galaxies \citep[e.g.][]{2013JCAP...03..014A} and a recent constraint from the high redshift UV luminosity function \citep{2016ApJ...825L...1M}. We discuss how our constraint will improve with a deeper stellar mass function and/or a complete census of low-mass isolated dwarfs in \S\ref{sec:discussion}.

\section{Discussion}\label{sec:discussion}

\subsection{A shallower group stellar mass function below $M_* \sim 10^9$\,M$_\odot$}

In \S\ref{sec:stellarmfuncs}, we argued that the stellar mass function is shallower in groups below $M_* \sim 10^9$\,M$_\odot$. It has been noted already in the literature that there are significant differences in both the luminosity and stellar mass functions of galaxy clusters and field galaxies \citep[e.g.][]{1998MNRAS.294..193T,2006ApJ...652..249X,2010A&A...524A..76B,2010ApJ...721..193P,2012ApJ...757....4P,2016arXiv160403957E}. However, a similar such environmental dependence on group scales has proven more elusive. Using SDSS data, \citet{2009ApJ...695..900Y} found no difference between the stellar mass function in groups or the field. However, they were only complete down to $M_* \sim 10^9$\,M$_\odot$ and so would not have been able to detect the difference that we find here. In principle, it should be possible to split the SDSS luminosity function in \citet{2005ApJ...631..208B} into a group and field sample to test our findings, but this is beyond the scope of this present work. As we noted in \S\ref{sec:stellarmfuncs}, it is compelling that \citet{2012ApJ...757...85G} report a field galaxy quenching mass scale of $M_* = 10^9$\,M$_\odot$ that depends on proximity to a larger host galaxy. This is precisely the stellar mass scale at which we calculated that ram pressure stripping will become important (\S\ref{sec:mquench}), and it is precisely the mass scale at which we find a suppression in the group stellar mass function. We will explore these ideas further in future work.

\subsection{The missing satellite problem and TBTF in groups and the field}\label{sec:missingsats_tbtf}

We have shown that abundance matching in $\Lambda$CDM is consistent with isolated dwarf galaxy rotation curves down to $M_{200} \sim 5 \times 10^9$\,M$_\odot$, and $M_{200} \sim 5 \times 10^8$\,M$_\odot$ if we assume a power law extrapolation of the SDSS stellar mass function. A direct corollary of this is that every single halo in the field is occupied with a galaxy down to these limits and, furthermore, that their gas dynamics are consistent with the halo that they live in. This means that there is {\it no `missing satellites' or TBTF problem in the field down to these limits.}

The above is interesting because both the missing satellites and TBTF problems occur in the Milky Way and Andromeda satellite population below a mass scale of $M_{\rm TBTF} \sim 10^{10}$\,M$_\odot$ \citep{2006MNRAS.tmp..153R,2011MNRAS.415L..40B,2014MNRAS.440.3511T}. If there is no similar problem at this mass scale for isolated `field' galaxies, then both problems must owe to some {\it environmental effect}. Indeed, a likely culprit is quenching due to ram pressure on infall to groups, as we have discussed already in \S\ref{sec:mquench}. It is beyond the scope of this paper to determine whether or not such environmental processes can fully solve the missing satellites and TBTF problems inside groups. However, it is hard to understand how some change to the underlying cosmological model could act inside groups but not in the field. For this reason, we assert that both of these small scale puzzles must owe to `galaxy formation physics', rather than exotic cosmology.

A final implication of the above result is that we expect significant scatter in $M_*$ for a given pre-infall $M_{200}$ inside groups. This means that, inside groups, classical `monotonic' abundance matching will fail \citep[see also][]{2015NatCo...6E7599U,2016arXiv160500004T}. However, more sophisticated mappings between dark and luminous subhalos that take account of the radial or orbit distribution of satellites could still work \citep[e.g.][]{2010MNRAS.402.1995M,2010arXiv1001.1731L,2013JCAP...03..014A}. Similarly, it may be possible to build a working abundance matching model that simply introduces significant scatter in the $M_*-M_{200}$ relation below some stellar mass scale \citep[e.g.][and see Appendix \ref{app:abundcompare}]{2016arXiv160304855G,2016arXiv161207834J}.

\subsection{Comparison with other works}\label{sec:comparison}

Our result that there is no missing satellites or TBTF problem for field dIrrs is apparently at odds with \citet{2015MNRAS.454.1798K} and \citet{2015A&A...574A.113P} who report a severe abundance/TBTF problem in the Local Volume. To arrive at this conclusion, both studies compare the distribution function of HI velocity line widths of a sample of Local Volume galaxies with predictions from the Bolshoi simulation. However, this relies on being able to convert HI velocity line widths to the peak rotation velocity of dark matter halos, $v_{\rm max}$. \citet{2016MNRAS.455.3841B} have argued that this conversion is complex, particularly for dwarfs with $v_{\rm max} < 50$\,km/s. With reasonable assumptions, they find that they can reconcile $\Lambda$CDM with the data in \citet{2015MNRAS.454.1798K} (and see also \citealt{2016A&A...591A..58P}). Recently, however, \citet{2016arXiv161009335T} have revived the debate. They take similar care with the conversion from HI line widths to $v_{\rm max}$, accounting for `cusp-core' transformations due to stellar feedback. Yet, they find that the Local Volume abundance problem persists. It is beyond the scope of this work to explore this further, but we note that if the stellar mass function is shallower inside groups, then it is likely to be suppressed also on the ${\sim}10$\,Mpc$^{3}$ scale of the Local Volume. If this is the case, then Local Volume galaxies should lie on the \MstarMrot\ relation that we find here, but have a stellar mass function that is shallower than SDSS.  

Our results are also in tension with the higher redshift study of \citet{2014ApJ...782..115M}. They derive an \MstarMrot\ relation for galaxies over the redshift range $0.2 < z < 1$, finding a significant offset from \MstarMabund. However, due to the higher redshift of their galaxy sample, they have only a single measure of the rotational velocity at 2.2 disc scale lengths. This is then extrapolated to the velocity at the virial radius $V_{200}$ via a weak lensing calibration at a stellar mass of $\log_{10}[M_*/M_\odot] = 9.0$. As highlighted by \citet{2014ApJ...782..115M}, this could introduce a potentially large systematic error. Furthermore, it is challenging with just a single measurement of the rotation velocity to identify `rogues' (see Figures \ref{fig:rotcurves_good}, \ref{fig:rotcurves_good2} and \ref{fig:rotcurves_bad}). We will explore the \citet{2014ApJ...782..115M} data further in future work.

More similar to our analysis here is the recent study of \citet{2016arXiv160505326P} (hereafter P16). They use the baryon-influenced mass models from \citet{2014MNRAS.441.2986D} to fit rotation curves and measure $M_{200}$ and $c$ for a large sample of dwarfs in the Little THINGS and THINGS surveys. Comparing their results with abundance matching predictions, similarly to our analysis here, they arrive at the opposite conclusion that $\Lambda$CDM is inconsistent with the data. Our analyses are sufficiently different that a detailed comparison is somewhat challenging, but we note here three key differences that likely lead to this apparent discrepancy: (i) P16 use the Little THINGS and THINGS rotation curves, whereas we derive the rotation curves using \Barolo\ \citep{2016arXiv161103865I}; (ii) P16 use the \citet{2014MNRAS.441.2986D} model that does not show cusp-core transformations below $M_{200} \sim 10^{10}$\,M$_\odot$, whereas we use the \coreNFW\ profile from R16a that does; and (iii) P16 primarily compare their results with \MstarMabund\ from \citet{2014MNRAS.444..222G}, whereas we favour matching the SDSS stellar mass function to the Bolshoi simulation. The most significant of these is (iii). The \citet{2014MNRAS.444..222G} \MstarMabund\ relation relies on {\it Local Group galaxies}. This is problematic since -- as we argued in \S\ref{sec:mstarmabund} -- classical abundance matching is expected to fail inside groups. Indeed, using the RT05 group stellar mass function, we derive an (erroneous) \MstarMabund\ relation that is remarkably similar to that derived in \citet{2014MNRAS.444..222G}, and a similar relation derived in \citealt{2014ApJ...784L..14B} (see Appendix \ref{app:abundcompare}). 

A similar critique explains the apparent discrepancy between our findings here and the earlier work of \citet{2012MNRAS.425.2817F}. They find, in agreement with us, that galaxies with stellar mass $M_* \simlt 3 \times 10^7$\,M$_\odot$ inhabit halos with mass $M_{200} \simlt 10^{10}$\,M$_\odot$. However, they argue that this is at odds with abundance matching in $\Lambda$CDM. Similarly to P16, this is because they use a steep \MstarMabund\ that is similar to our erroneous RT05 `group' relation. Using the shallower field galaxy SDSS \MstarMabund, the \citet{2012MNRAS.425.2817F} results are in good agreement with ours.

Finally, \citet{2016arXiv160505971K} have recently compared the \citet{2014MNRAS.441.2986D} model to data for 147 rotation curves from the SPARC sample. Comparing their derived \MstarMrot\ relation with \MstarMabund, they conclude similarly to us here that $\Lambda$CDM works very well. In this case, there is no discrepancy. \citet{2016arXiv160505971K} focus on galaxies that are substantially more massive than those we study here, with $M_{200} > 10^{10}$\,M$_\odot$. In this sense, the \citet{2016arXiv160505971K} study is wholly complementary to ours that focusses on the regime $M_{200} < 10^{10}$\,M$_\odot$.

\subsection{The interesting outlier DDO 154}\label{sec:ddo154}

As discussed in \S\ref{sec:mstarmrot}, our \MstarMrot\ relation in Figure \ref{fig:carina_plot} has one significant outlier, DDO 154. This galaxy also has an unusually high HI gas mass fraction, with $M_{\rm HI}/M_* = 37$. At its currently observed star formation rate of $\dot{M}_* = 3.82 \times 10^{-3}$ \citep{2012AJ....143...47Z}, DDO 154 would move onto our \MstarMrot\ relation in ${\sim}5.7$\,Gyrs. This is an interesting timescale. In $\Lambda$CDM, most major galaxy mergers are complete by redshift $z=1$ some ${\sim}8$\,Gyrs ago \citep[e.g.][]{2009ApJ...702.1005S}. Thus, if post-merger isolated dwarfs look like DDO 154, then most would have had time to deplete their excess HI gas and move onto the \MstarMrot\ relation by today. A possible explanation for DDO 154, then, is that it has just undergone a relatively rare late merger. We will explore this idea further in future work. 

\subsection{How close is too close?}\label{sec:carina}

It is interesting to ask {\it how close} to the Milky Way satellites can orbit before they become quenched. The Carina dwarf spheroidal is particularly interesting in this regard. Its orbit remains highly uncertain due to its large proper motion errors, but \citet{2010arXiv1001.1731L} find that it seems to be substantially more circular than the mean of subhalos in a $\Lambda$CDM pure dark matter simulation. With an apo-to-pericentre ratio of $r_p/r_a = 0.3-0.7$, it is also potentially more circular than all of the other Milky Way dwarfs, except one: Fornax. Fornax is on a cosmologically unusual near circular orbit, with $r_p/r_a \sim 0.6-0.8$  \citep{2010arXiv1001.1731L}. Along with Carina, it is the only other Milky Way dwarf spheroidal that has continued to form stars for nearly a Hubble time \citep{2013MNRAS.433.1505D}. Such circularity may be the key to these dwarfs' ability to continue to form stars, lending further support to the idea that quenching is driven primarily by ram pressure. From equation \ref{eqn:rampressure}, we can see that ram pressure is proportional to the satellite velocity at pericentre squared: $v_p^2$. Circular orbits minimise $v_p$ and will therefore also minimise the effect of ram pressure stripping. The fact that Carina appears to lie on the \MstarMrot\ relation of isolated dwarfs suggests that it has come just about as close to the Milky Way as possible while maintaining its ability to form stars. This may help to explain its puzzlingly unique star formation history \citep{2014A&A...572A..10D}.

\subsection{Implications for cusp-core transformations at low stellar mass}\label{sec:dmcoresalltheway}

In R16a, we found that dark matter cusp-core transformations for isolated dwarfs continue down to at least $M_* \sim 5 \times 10^5$\,M$_\odot$ ($M_{200} \sim 5\times 10^8$\,M$_\odot$), under the assumption that reionisation does not shut down star formation at this mass scale (see \S\ref{sec:intro} and \S\ref{sec:reionisation} for more discussion on this point). However, several works in the literature have claimed that there is insufficient energy in such low stellar mass systems for cusp-core transformations to proceed \citep[e.g.][]{2012ApJ...759L..42P,2013MNRAS.433.3539G,2014MNRAS.441.2986D,2015arXiv150703590T}. We are now in a position to revisit this problem. As discussed in R16a, the main difference between all of the studies in the literature to date has been in the stellar mass to halo mass relation (either assumed or self-consistently calculated using hydrodynamic simulations). The more stars a given halo forms, the more supernovae it has to unbind its dark cusp.

Following \citet{2012ApJ...759L..42P} and R16a, we may estimate the supernova energy that is available to  transform cusps to cores as:

\begin{equation} 
\Delta E = \frac{E_{\rm SN} M_*}{\langle m_* \rangle} \xi(m_* > 8\,{\rm M}_\odot) \epsilon_{\rm DM}
\label{eqn:pencalc}
\end{equation}
where $E_{\rm SN} = 10^{51}$\,erg is the energy of a single supernova; $\langle m_* \rangle = 0.83$\,M$_\odot$ is the mean stellar mass; $\xi = 0.00978$ is the fraction of mass in stars that go supernova\footnote{As in R16a, we assume a \cite{chabrier03} IMF averaged over the range $0.1 < m_*/{\rm M}_\odot < 100$.}; and $\epsilon_{\rm DM} \simeq 0.25 - 0.8\%$ is the efficiency of coupling of the SNe energy to the dark matter. (We estimate $\epsilon_{\rm DM}$ using the simulations in R16a. Following R16a, this is defined as the ratio of the energy required to unbind the dark matter cusp to the integrated supernovae energy.)

The available supernova energy can then be compared with the energy required to unbind the dark matter cusp:

\begin{equation} 
\Delta W = -\frac{1}{2}\int_0^\infty \frac{G(M_{\rm NFW}^2 - M_{\rm cNFW}^2)}{r^2} dr
\label{eqn:deltaW}
\end{equation}
where $M_{\rm NFW}$ and $M_{\rm cNFW}$ are the enclosed cumulative mass for an \NFW\ and \coreNFW\ dark matter density profile, respectively (equations \ref{eqn:MNFW} and \ref{eqn:coreNFW}).

\begin{figure}
\begin{center}
\includegraphics[width=0.47\textwidth]{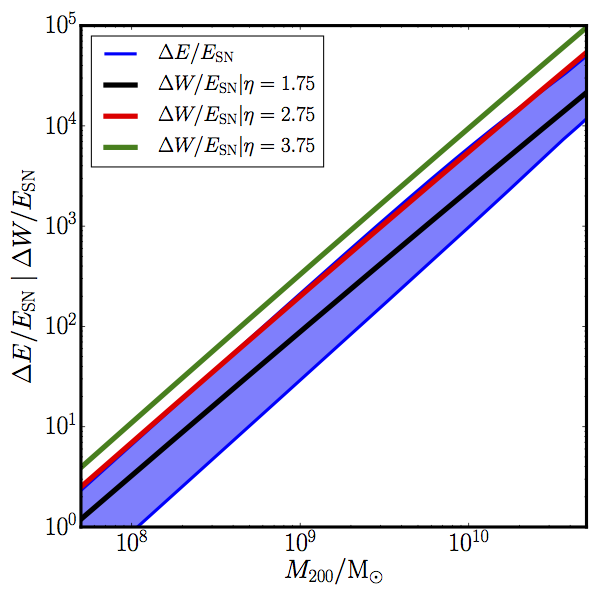}
\caption{The supernova energy available for driving dark matter cusp-core transformations (blue band) compared to the energy required to unbind the dark matter cusp (black, red and green lines), as a function of the dark matter halo mass $M_{200}$.  The black line shows results for our default dark matter core size of $\eta = 1.75$ (equation \ref{eqn:etarc}). The red and green lines show results for larger dark matter cores with $\eta = 2.75$ and $\eta = 3.75$, respectively. The energies are plotted in units of a single supernova explosion ($E_{\rm SN}$). For our default dark matter core size ($\eta = 1.75$) there is sufficient energy from SNe explosions at all mass scales to excite cusp-core transformations `all the way down'.}
\label{fig:dmcores_alltheway} 
\end{center}
\end{figure}

The available energy (equation \ref{eqn:pencalc}) depends on the galaxy stellar mass $M_*$, while the required energy to unbind the cusp (equation \ref{eqn:deltaW}) depends on the halo mass $M_{200}$. Hence, the $M_*-M_{200}$ relation is critical. Using the \MstarMabund\ relation from Figure \ref{fig:carina_plot}, we plot $\Delta E(M_{200})$ and $\Delta W(M_{200})$ in Figure \ref{fig:dmcores_alltheway}. We assume that the stellar half mass radius is given by $R_{1/2} \sim r_{1/2} \sim 0.015 r_{200}$ \citep{2013ApJ...764L..31K,2015ApJS..219...15S,2015arXiv150900853A}; that the total star formation time $t_{\rm SF} = 14$\,Gyrs such that core formation is complete; that halos obey the concentration mass relation from \citet{2007MNRAS.378...55M}; and that the birth stellar mass (i.e. before mass loss due to stellar evolution) is ${\sim}2$ times the current stellar mass (see R16a). 

The results are shown in Figure \ref{fig:dmcores_alltheway}, where the blue band marks the available supernova energy as a function of halo mass $M_{200}$, while the black, red and green lines mark the energy required to unbind the cusp. The black line shows results for our default dark matter core size of $\eta = 1.75$ (equation \ref{eqn:etarc}). The red and green lines show results for larger dark matter cores with $\eta = 2.75$ and $\eta = 3.75$, respectively. As can be seen, there is sufficient energy to unbind a dark matter cusp `all the way down' for $\eta < 2.75$, but insufficient energy to build larger cores than this. 

These results support our assertion in R16a that dark matter cores can form `all the way down'. However, this is only energetically possible if isolated low mass halos are largely unaffected by reionisation. We discuss this, next.

\subsection{But what about reionisation?}\label{sec:reionisation}

Our results suggest that all field dark matter halos are occupied with galaxies down to $M_{200} \sim 5 \times 10^8$\,M$_\odot$ (if we assume a power law extrapolation of the SDSS stellar mass function). Indeed, Leo T -- which appears to sit at this mass scale -- has formed stars continuously at a rate of just $\sim 10^{-5}$\,M$_\odot$\,yr$^{-1}$ for a Hubble time \citep{2012ApJ...748...88W}. If Leo T does inhabit such a low mass halo, then a corollary of this is that reionisation does not appear to suppress galaxy formation above $M_{\rm reion} \sim 5 \times 10^8$\,M$_\odot$. This is in excellent agreement with recent models by \citet{2014ApJ...793...30G}, but at tension with other simulations that favour a substantially higher $M_{\rm reion} \simgt 3 \times 10^9$\,M$_\odot$ \citep[e.g.][]{2013MNRAS.432.1989S,2015arXiv150402466W,2015arXiv151100011O,2016MNRAS.456...85S,2016arXiv161102281F}.

The above tension is not necessarily a cause for concern. There is a least a factor $2-4$ uncertainty in the flux of ionising photons at high redshift, with galaxies being the dominant ionising source at $z\gtrsim 3$ and quasars dominating at lower redshifts \citep[e.g.][]{Haardt2012}. Large volume simulations are required to capture these photon sources correctly. These require very high spatial resolution and accurate ray propagation to model self shielding effects and to correctly predict the photon escape fraction, $f_{\rm esc}$ (e.g. \citealt{2016arXiv160105802G}). In particular, \cite{KimmCen2014} found, using high resolution simulations of galaxies forming in haloes with virial masses ${\sim}10^8-10^{10}$M$_\odot$ at $z\gtrsim 7$, that $f_{\rm esc}$ fluctuates by orders of magnitude over a dynamical time due to stellar feedback \citep[and see also][]{Trebitsch2015}. Numerical resolution is also important. \cite{Hawthorn2015} find that reionisation blows out far more gas in low resolution simulations as compared to higher resolution simulations that better-capture dense gas. Finally, it is important to include all of the important physics. \citet{2009MNRAS.392L..45R} highlight the importance of adiabatic cooling due to the expansion of the Universe; while \citet{2016MNRAS.458..912V} suggest that additional feedback due to Population III stars could reduce early star formation and feedback, leading to less hot diffuse gas and less reionisation-driven gas blow out. It is beyond the scope of this present work to explore these ideas in more detail. We simply note here that at present there is no cause for concern if Leo T inhabits a dark matter halo of mass $M_{200} \sim 5 \times 10^8$\,M$_\odot$.

\subsection{Implications for near-field cosmology} 

Our results allow us to make several concrete predictions for upcoming near-field cosmology surveys:

\begin{itemize}

\item Firstly, assuming that $\Lambda$CDM is correct, we predict that the stellar mass function of field galaxies should continue as an unbroken power law with slope $\alpha \sim 1.6$, at least over the mass range $10^5 < M_*/{\rm M}_\odot < 10^7$. Testing this will require large volume surveys like SDSS to avoid contamination from groups.

\item Secondly, below $M_* \sim 10^5$\,M$_\odot$, we may see the first signs of star formation truncation due to reionisation. The smoking gun for this would be an extremely isolated quenched dwarf. However, as discussed in \citet{2012ApJ...757...85G}, to be classified as `extremely isolated' it would need to be found ${\simgt}4$ virial radii away from any nearby larger galaxy. 

\item Thirdly, our results imply that there should be many galaxies like Leo T on the outskirts of the Milky Way and Andromeda just waiting to be found. Extrapolating the Bolshoi mass function to low mass, we predict that there should be ${\sim}2000$ galaxies like Leo T in a typical 10\,Mpc$^3$ volume, with halo mass $5 \times 10^8 < M_{200}/{\rm M}_\odot < 10^9$; stellar mass ${\sim}2\times 10^5 < M_*/{\rm M}_\odot < 6 \times 10^5$; and HI gas mass ${\sim}3 \times 10^5 < M_{\rm HI}/{\rm M}_\odot < 3 \times 10^6$. In practice, this will be an upper bound because many of these `Leo T'-like dIrrs will have been ram pressure stripped by a nearby host galaxy. Nonetheless, it is tantalising that Leo T lies right on the edge of the SDSS survey footprint \citep{2009ApJ...696.2179K}. Any closer to the Milky Way and Leo T would have been stripped of its gas, similarly to the recently discovered Eridanus II galaxy \citep{2016ApJ...824L..14C}. Any further away, and it would have been too faint to be seen. Indeed, \citet{2016arXiv161105888J} have recently discovered a slew of new star forming dIrrs in the Local Volume. These may be the tip of the iceberg.

\end{itemize}

Finally, we note that the comparison between \MstarMabund\ and \MstarMrot\ shows great promise for constraining $m_{\rm WDM}$ if we can reach down to Leo T mass galaxies and below. This suggests that it is worth the effort of attempting to model the Local Group at the fidelity of the simulations presented in R16a, despite the computational challenges that this presents. At least some of the `ultra-faint' dwarfs that have already been found orbiting the Milky Way and Andromeda are likely even less massive than Leo T \citep[e.g.][]{2013ApJ...770...16K}, holding the promise of providing unparalleled constraints on $m_{\rm WDM}$ and/or other cosmologies that suppress small scale power.

\section{Conclusions}\label{sec:conclusions}

We have presented a clean probe of cosmology on small scales that follows from the comparison of \MstarMrot, measured from the rotation curves of isolated dwarf galaxies in the field, and \MstarMabund\ calculated from abundance matching (see \S\ref{sec:newprobe}). These should agree if the cosmological model is correct, but will diverge if the halo mass function is too shallow or steep on small scales. Our probe is comparatively clean since it relies only on the following theory ingredients: (i) a monotonic relation between stellar mass and halo mass; (ii) a predicted dark matter halo mass function; and (iii) a robust prediction of the internal dark matter distribution in dwarf irregular galaxies, for a given cosmological model. The first of these can be empirically tested using \MstarMrot; while (ii) and (iii) are readily obtained from state-of-the art numerical simulations (see \S\ref{sec:newprobe}).

Our key results are as follows: 

\begin{itemize}

\item We fit the rotation curves of a carefully selected sample of 19 isolated dIrr galaxies. Of these, five were found to be of too low inclination to be reliably inclination corrected (`inclination rogues'); another two (DDO 216 and NGC 1569) showed clear signs of disequilibrium (`disequilibrium rogues'); while one (DDO 101) had a very large distance uncertainty (`distance rogues'). For the remaining 11 dIrrs, we found that an \NFW\ dark matter halo profile is ruled out at $>99$\% confidence, reaffirming the well known `cusp-core' problem. By contrast, the \coreNFW\ profile from R16a -- that accounts for cusp-core transformations due to stellar feedback -- gives an excellent fit in all cases, without introducing any more free parameters than the \NFW\ form.

\item Although we required the \coreNFW\ profile to obtain a good fit to the rotation curve shape, we showed that the implied dark matter halo mass $M_{200}$ was not sensitive to the form of the dark matter density profile within $r \simlt R_{1/2}$. For this reason, we were able to robustly measure the stellar mass-halo mass relation \MstarMrot\ over the mass range $5 \times 10^5 \simlt M_{*}/{\rm M}_\odot \simlt 10^{8}$, finding a monotonic relation with little scatter.

\item Such monotonicity implies that abundance matching should yield a \MstarMabund\ relation that matches \MstarMrot, if the cosmological model is correct. Using the `field galaxy' stellar mass function from the Sloan Digital Sky Survey (SDSS) and the halo mass function from the $\Lambda$CDM Bolshoi simulation, we found remarkable agreement between the two. This held down to $M_{200} \sim 5 \times 10^9$\,M$_\odot$, and to $M_{200} \sim 5 \times 10^8$\,M$_\odot$ if we assumed a power law extrapolation of the SDSS stellar mass function below $M_* \sim 10^7$\,M$_\odot$. 

\item The good agreement between \MstarMrot\ and \MstarMabund\ means that there is no `missing satellites' or TBTF problem for our sample of isolated dIrrs down to at least $M_{200} \sim 5 \times 10^9$\,M$_\odot$. This is lower than the mass scale at which the `missing satellites' and TBTF problems manifest in the Local Group, $M_{\rm TBTF} \sim 10^{10}$\,M$_\odot$ \citep[e.g.][]{2006MNRAS.tmp..153R,2011MNRAS.415L..40B,2014MNRAS.440.3511T}. This suggests that both problems depend on {\it environment} and therefore owe to `galaxy formation physics' rather than exotic cosmology.

\item Compiling stellar mass functions from the literature, we showed that the group stellar mass function is substantially shallower than the field below $M_* \sim 10^9$\,M$_\odot$. We argued that this likely owes to ram pressure stripping on group infall. This induces a significant scatter in $M_*$ for a given pre-infall $M_{200}$ causing classical abundance matching to fail.

\item We considered how well a $\Lambda$ Warm Dark Matter ($\Lambda$WDM) cosmology can fit \MstarMrot. Repeating our abundance matching using the SDSS field stellar mass function, we showed that $\Lambda$WDM fails at 68\% confidence for a thermal relic mass of $m_{\rm WDM} < 1.25$\,keV, and $m_{\rm WDM} < 2$\,keV if we used the power law extrapolation of the SDSS stellar mass function.

\item If $\Lambda$CDM is correct, we predict that the stellar mass function of galaxies should continue as an unbroken power law with slope $\alpha \sim 1.6$, at least over the mass range $10^5 < M_*/{\rm M}_\odot < 10^7$. There should be ${\sim}2000$ galaxies like Leo T in a typical 10\,Mpc$^3$ volume, with halo mass $5 \times 10^8 < M_{200}/{\rm M}_\odot < 10^9$; stellar mass ${\sim}2\times 10^5 < M_*/{\rm M}_\odot < 6 \times 10^5$; and HI gas mass ${\sim}3 \times 10^5 < M_{\rm HI}/{\rm M}_\odot < 3 \times 10^6$. Below this mass scale, we may see the first signs of star formation suppression due to reionisation.

\end{itemize} 

\section{Acknowledgements}

JIR and OA would like to acknowledge support from STFC consolidated grant ST/M000990/1. JIR acknowledges support from the MERAC foundation. OA would like to acknowledge support from the Swedish Research Council (grant 2014-5791). This research made use of APLpy, an open-source plotting package for Python hosted at {{\tt http://aplpy.github.com}}. This work used PyNbody for the simulation analysis ({{\tt https://github.com/pynbody/pynbody}}; \citealt{2013ascl.soft05002P}). All simulations were run on the Surrey Galaxy Factory. We would like to thank Se-Heon Oh and Little THINGS for kindly making their data public, and Erwin de Blok for providing the data for NGC 6822. We would like to thank Andrew Pontzen, Julio Navarro, Kyle Oman, Peter Behroozi and Annika Peter for useful discussions. We would like to thank the anonymous referee for useful comments that improved the clarity of this work.

\appendix 

\section{The rotation curve fits}\label{app:rotcurves}

In this Appendix, we show the rotation curves and model fits for all of the galaxies listed in Table \ref{tab:data}. In Figures \ref{fig:rotcurves_good} and \ref{fig:rotcurves_good2}, we show all of the galaxies that we include in Figure \ref{fig:carina_plot}; in Figure \ref{fig:rotcurves_bad}, we show all of the `rogues' that we exclude from further analysis. The rogues fall into three categories: `inclination' ($i$-Rogues); `disequilibrium' (Diseq. Rogues); and `distance' ($D$-Rogues), as marked (see \S\ref{sec:rogues}). The first class of these three have uncertain inclination, with $i_{\rm fit} < 40^\circ$. The second class show signs of disequilibrium, either in the form of significant fast-expanding HI bubbles (see Table \ref{tab:data}), or -- as is the case for Pegasus -- because we only have data for the inner rotation curve inside $R_{1/2}$. This inner region is particularly susceptible to disequilibrium effects from both supernova-driven HI holes and non-circular motions. (This can be seen, for example, in the inner rotation curves of NGC 6822 and DDO 126 that are otherwise both in our `clean' sample.) The third class (which contains only DDO 101) have very uncertain distance $D$ (see R16b for a detailed discussion of this galaxy). The individual galaxies are discussed in more detail in \citet{2016arXiv161103865I} where we present the full details of our rotation curve derivation, including a comparison with the rotation curves from Little THINGS.

\begin{figure*}
\begin{center}
\includegraphics[width=0.99\textwidth]{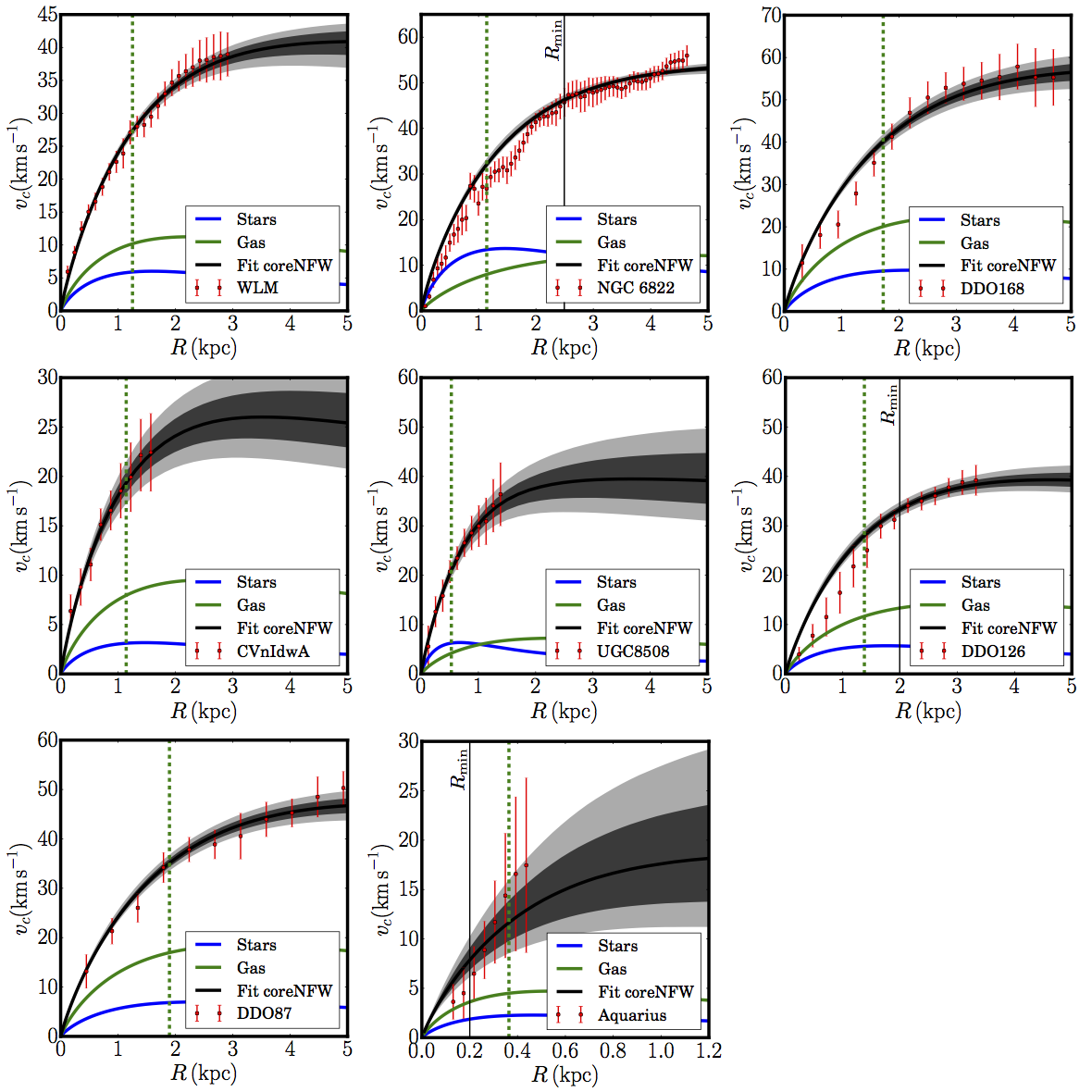}
\caption{Rotation curve data (red data points) and models for our sample of `clean' dIrr galaxies (see Table \ref{tab:data}). The black contours show the median (black), 68\% (dark grey) and 95\% (light grey) confidence intervals of our fitted \coreNFW\ rotation curve models (see \S\ref{sec:massmodel}). The vertical green dashed line shows the projected stellar half light radius $R_{1/2}$. The thin vertical black line marks the inner data point used for the fit, $R_{\rm min}$ (where this is not marked $R_{\rm min} = 0$). The blue and green lines show the rotation curve contribution from stars and gas, respectively.} 
\label{fig:rotcurves_good} 
\end{center}
\end{figure*}

\begin{figure*}
\begin{center}
\includegraphics[width=0.99\textwidth]{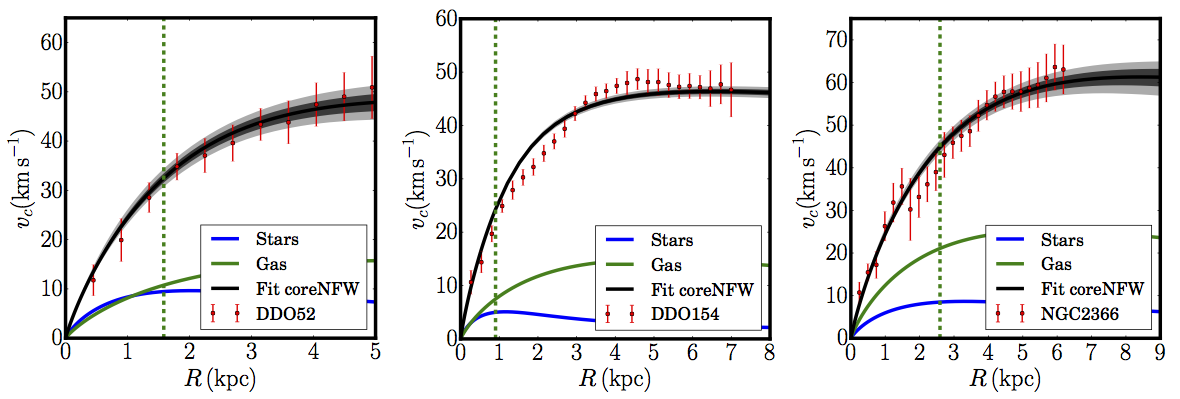}
\caption{Continuation of Figure \ref{fig:rotcurves_good}.} 
\label{fig:rotcurves_good2} 
\end{center}
\end{figure*}

\begin{figure*}
\begin{center}
\includegraphics[width=0.99\textwidth]{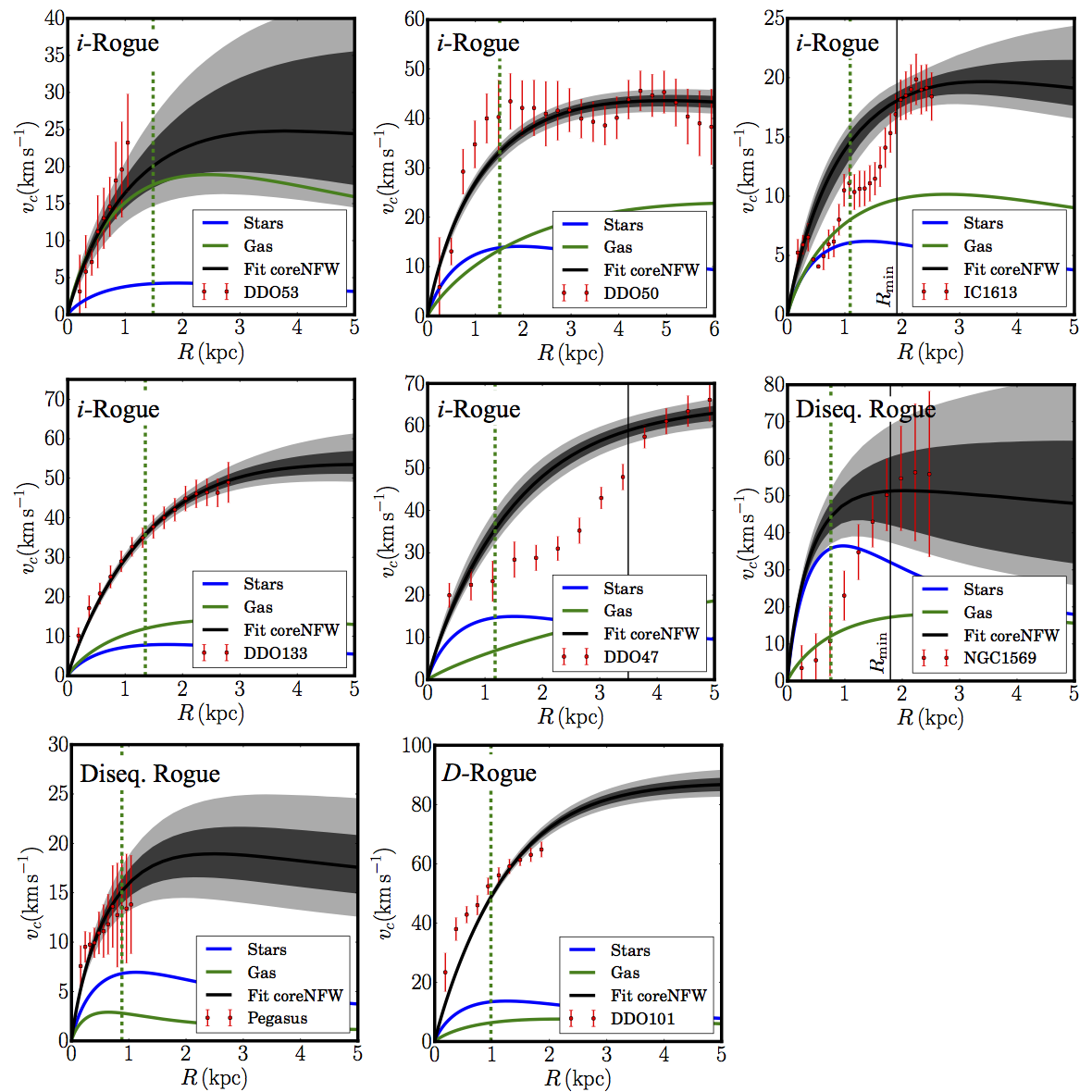}
\caption{Rotation curve data  and models for our `rogue' dIrr galaxies (see Table \ref{tab:data}). The lines and symbols are as in Figure \ref{fig:rotcurves_good}. The rogue galaxies fall into three categories: `inclination' ($i$-Rogues); `disequilibrium' (Diseq. Rogues); and `distance' ($D$-Rogues), as marked.}
\label{fig:rotcurves_bad} 
\end{center}
\end{figure*}

\section{Testing the robustness of our \MstarMrot\ relation}\label{app:robust}

\begin{figure*}
\begin{center}
\includegraphics[width=0.99\textwidth]{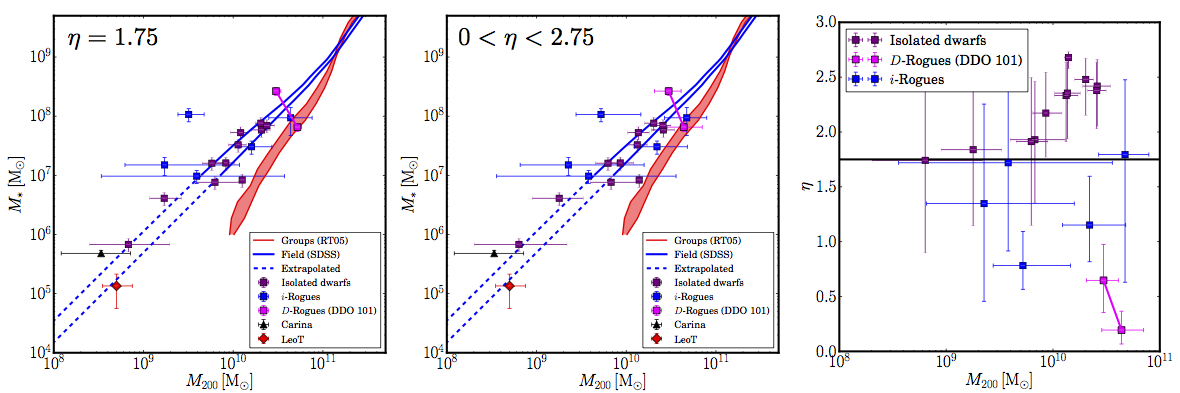}
\caption{Testing the robustness of our \MstarMrot\ relation. In the left panel, we show \MstarMrot\ including the $i$-Rogues (blue data points) and $D$-Rogue (magenta data points); the lines and symbols are as in Figure \ref{fig:carina_plot}. In the middle panel, we show the same but now marginalising over the dark matter core size, using a flat prior on $0 < \eta < 2.75$ (see equation \ref{eqn:etarc}). In the right panel, we show the marginalised $\eta$ parameters for each galaxy that result from this fit, including the $i$-Rogues (blue data points) and $D$-Rogue (magenta data points). The horizontal black line marks our default $\eta = 1.75$.}
\label{fig:robust} 
\end{center}
\end{figure*}

In this Appendix, we explore how robust our \MstarMrot\ relation is to key assumptions in our methodology. In Figure \ref{fig:robust}, left panel, we show our \MstarMrot\ relation including the $i$-Rogues (blue data points) and $D$-Rogue (magenta data points). Recall that for the $i$-Rogues, we marginalise over the inclination angle with a flat prior over the range $0^\circ < i < 40^\circ$ (see \S\ref{sec:rotmethod}). In most cases, this significantly inflates the uncertainties. However, DDO 133 has sufficiently good data that -- under the assumption of a \coreNFW\ dark matter profile -- its rotation curve shape is sufficient to provide a measurement of $i$ (see Figure \ref{fig:rotcurves_bad} and Table \ref{tab:data}). 

As can be seen in Figure \ref{fig:robust}, left panel, including the $i$-Rogues introduces substantially more scatter about the \MstarMrot\ relation, but no more than is expected given their larger uncertainties on $M_{200}$. Thus, the $i$-Rogues do not substantially alter our key results and conclusions. We do, however, find an additional outlier, DDO 50, that appears to have a very high $M_*$ for its $M_{200}$, even when marginalising over $i$. As can be seen in Figure \ref{fig:rotcurves_bad}, however, DDO 50 has an unusual rotation curve with prominent wiggles out to large radii. This makes it challenging to inclination correct. Indeed, from a stability analysis, \citet{2014RMxAA..50..225S} argue for an inclination for DDO 50 of $i = 27^\circ$ that is substantially smaller than the ${\sim}37^\circ$ that we find here (see Table \ref{tab:data}). This lower inclination would be sufficient to push DDO 50 onto the \MstarMrot\ relation.

Our one $D$-Rogue -- DDO 101 -- is marked on Figure \ref{fig:robust}, left panel, by the magenta data points. Since the distance for this galaxy is highly uncertain (see the discussion in R16b), we plot two points for $D_{\rm DDO101} = 6.4$\,Mpc and $D_{\rm DDO101} = 12.9$\,Mpc. As can be seen, these straddle the \MstarMrot\ relation. Thus, our $D$-Rogue does not affect our key results and conclusions either. 

Finally, we consider the effect of allowing the dark matter core size to freely vary by performing our rotation curve fits assuming a flat prior on $\eta$ (equation \ref{eqn:etarc}) over the range $0 < \eta < 2.75$. (The upper bound on $\eta$ is set by the energetic arguments in \S\ref{sec:dmcoresalltheway}. There, we showed that the integrated supernova energy is not sufficient to build cores larger than ${\sim}2.75 R_{1/2}$; see Figure \ref{fig:dmcores_alltheway}.) The results for \MstarMrot\ are shown in Figure \ref{fig:robust}, middle panel; while Figure \ref{fig:robust}, right panel, shows the marginalised $\eta$ parameters for each galaxy that result from this fit, including the $i$-Rogues (blue data points) and $D$-Rogue (magenta data points). The horizontal black line marks our default $\eta = 1.75$. 

From Figure \ref{fig:robust}, middle panel, we see that allowing $\eta$ to vary increases our errors on $M_{200}$ but does not otherwise affect our key results or conclusions. This is consistent with our findings in \S\ref{sec:exrots}, where we showed that fitting an \NFW\ profile to the rotation curves gives a poorer fit, but does not significantly alter the derived $M_{200}$ within our quoted uncertainties. Finally, from the rightmost panel of Figure \ref{fig:robust}, we can see that we do not obtain very strong constraints on $\eta$, similarly to our findings for WLM in R16b. The data are consistent with our default $\eta = 1.75$, with perhaps a hint that the more massive galaxies (with $M_{200} \simgt 10^{10}$\,M$_\odot$) favour a slightly larger core. For our sample of `clean' dIrrs (purple data points), we can definitively rule out a cusp ($\eta = 0$) at greater than $99$\% confidence.

\section{Comparison with other abundance matching work in the literature}\label{app:abundcompare}

In this Appendix, we compare our abundance matching curves in Figure \ref{fig:carina_plot} with other determinations in the literature from \citet{2010ApJ...710..903M}; \citet{2014ApJ...784L..14B}; and \citet{2014MNRAS.444..222G} (G-K14). \citet{2010ApJ...710..903M} perform a parametric `classical' abundance matching of SDSS galaxies in $\Lambda$CDM down to $M_{200} \sim 3 \times 10^{10}$\,M$_\odot$. Below this mass scale, the extrapolation of the \citet{2010ApJ...710..903M} relation diverges from our abundance matching curve shown in blue. However, this extrapolation is not supported by the latest data from SDSS. Over the mass range $7 \times 10^9 < M_{200} / {\rm M}_\odot < 3 \times 10^{10}$, our relation (that is based on deeper SDSS data than \citealt{2010ApJ...710..903M}; see \citealt{2010ApJ...717..379B,2013ApJ...770...57B}) diverges significantly from the \citet{2010ApJ...710..903M} extrapolation, suggesting that the \citet{2010ApJ...710..903M} relation should not be used below $M_{200} \sim 3 \times 10^{10}$\,M$_\odot$.

\citet{2014ApJ...784L..14B} (black data points) and \citet{2014MNRAS.444..222G} (G-K14; solid black line) reach to much lower stellar mass than SDSS by using constrained simulations of the Local Volume in $\Lambda$CDM abundance-matched to Local Group galaxies. In this sense, they are both similar to abundance matching with the group stellar mass function of RT05. In particular, both studies -- like our RT05 analysis -- rely on the assumption that the Local Group satellites have a monotonic relation between stellar mass and halo mass. Thus, it is perhaps not surprising that \citet{2014ApJ...784L..14B}, \citet{2014MNRAS.444..222G} and RT05 all agree within their 95\% confidence intervals. All three are substantially steeper than the SDSS abundance matching curve (blue). We argued in \S\ref{sec:mstarmabund} and \S\ref{sec:missingsats_tbtf} that the assumption of monotonicity is expected to break down inside groups \citep[see also][]{2015NatCo...6E7599U}. Indeed, the poor correspondence between the group abundance matching \MstarMabund\ and \MstarMrot\ is evidence for this. Thus, we conclude that, similarly to our RT05 abundance matching relation, the G-K14 and \citet{2014ApJ...784L..14B} relations are likely flawed due to the erroneous assumption of monotonicity.

Finally, \citet{2016arXiv160304855G} explore relaxing the monotonicity assumption for Local Group galaxies by adding significant scatter to the stellar mass-halo mass relation below some stellar mass scale. They show that this causes \MstarMabund\ to {\it steepen}, consistent with the apparently steeper group \MstarMabund\ relation that we find here. Adding such scatter to classical abundance matching is a promising avenue for probing $\Lambda$CDM in group environments, as has been demonstrated recently by \citet{2016arXiv161207834J}.

\begin{figure}
\begin{center}
\includegraphics[width=0.49\textwidth]{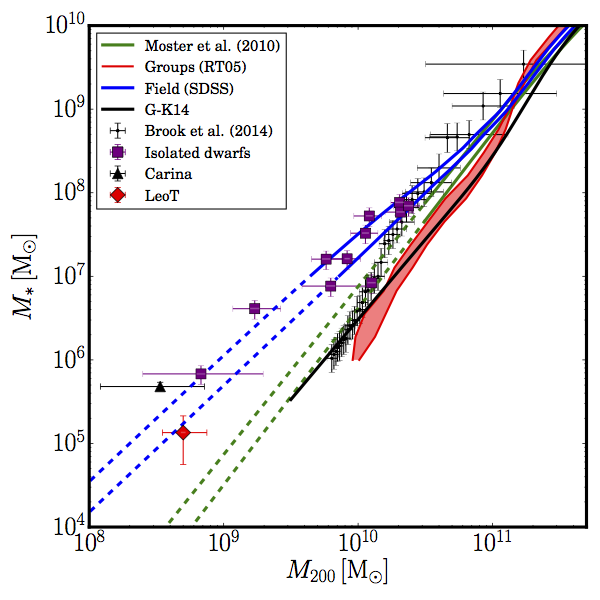}
\caption{A comparison of different abundance matching curves from the literature. The lines and symbols are as in Figure \ref{fig:carina_plot}, but we include abundance matching curves from \citet{2010ApJ...710..903M} (green); \citet{2014ApJ...784L..14B} (black data points); and \citet{2014MNRAS.444..222G} (G-K14; black solid line).}
\label{fig:carina_plot_compare} 
\end{center}
\end{figure}

\bibliographystyle{mnras}
\bibliography{./refs,ref2}

\end{document}